\documentclass[11pt]{article}

\usepackage{jheppub}
\usepackage[utf8]{inputenc}
\usepackage[T1]{fontenc}
\usepackage{amsmath, amsfonts, amssymb, amsthm, mathtools, mathrsfs, dsfont}
\usepackage{physics, tensor}
\usepackage{enumitem}
\usepackage{array}
\usepackage[dvipsnames]{xcolor}

\usepackage{graphicx}

\usepackage{parskip}
\usepackage{tikz}
\usetikzlibrary{arrows.meta, decorations.pathmorphing, math, calc, decorations.markings, positioning}

\title{Wald-like entropy and Islands in Dimensionally Reduced Einstein-Hilbert Gravity}
\author[a, b]{Krishna Jalan}
\affiliation[a]{The Institute of Mathematical Sciences, IV Cross Road, C.I.T. Campus, Taramani, Chennai, India 600113}
\affiliation[b]{Homi Bhabha National Institute, Training School Complex, Anushakti Nagar, Mumbai, India 400094}
\emailAdd{krishnajalan@imsc.res.in}

\abstract{We study the island formula and Page curve for the asymptotically flat eternal and quasi-stationary evaporating black hole solutions within the dimensionally reduced Einstein-Hilbert (DREH) model. In this model, the four-dimensional Einstein-Hilbert action reduces to a two-dimensional dilaton gravity, on top of which the quantum corrections can be incorporated via the Polyakov-Liouville (PL) action for a two-dimensional conformal field theory with a large central charge $c$. This gravity model arises from the s-wave approximation of four-dimensional Einstein-Hilbert gravity and provides a fully gravitational setting in which we study the islands, i.e., we will not require a non-gravitational bath region to collect the radiation, instead it lives in the black hole spacetime itself. The fine-grained entropy of Hawking radiation in the eternal and evaporating black holes within the DREH model was derived using the island rule in \cite{djordjevic2022eternal, djordevic2025evaporating}. The island rule is based on the replica method using the Euclidean gravitational path integral. In this work, we complement the Euclidean approach by providing a Lorentzian prescription for the generalized entropy $S_{\text{gen}}$, in the island formula without invoking the replica trick to compute $S_{\text{gen}}$. The generalized entropy is shown to coincide with Wald-like Noether charge of the combined DREH-PL action. Using this generalized entropy, we determine the quantum extremal surface, the Page time, and the Page curve for the eternal and the evaporating black hole. We conclude with a discussion of the possible extension to the full four-dimensional geometry incorporating vacuum polarization corrections. \vspace{30pt}}

\dedicated{Dedicated to the memory of A. P. Balachandran, mentor and friend.}

\begin{document}
	\maketitle

	\newcommand{\scri}{\mathscr{I}}
	\newcommand{\T}{T^{(\vec{\chi})}}
	\newcommand{\normT}[2][ab]{:\!T_{#1}(#2)\!:}
	\newcommand{\normTchi}[2][ab]{:\!T^{(\vec{\chi})}_{#1}(#2)\!:}
	\newcommand{\pd}{\partial}

	\section{Introduction} \label{sec:intro}
	Developments in the 1970s resulted in the conclusion that black holes are thermodynamics objects with entropy and temperature \cite{bekenstein1972second, bekenstein1973BH, bardeen1973BHT}. It also opened up several avenues for exploring the quantum aspects of gravity with black holes at the focal point. Perhaps the most prominent development was Hawking's famous discovery of the black hole radiance \cite{hawking1974BH, hawking1975particle} which also lead to the black hole information puzzle \cite{hawking1976breakdown}. The information puzzle states that the process of black hole formation and eventual evaporation violates the quantum mechanical unitary evolution. This is because the emitted radiation only depends on the few black hole charges which do not capture the complete information about the initial state of the black hole, leading to a loss of information. Since Hawking's work there have been several refinements of the original puzzle, \cite{mathur2009BHIP, page1993information, page1994BHI, almheiri2012AMPS, page2013HawkRad, almheiri2013AMPSS}. For a review on recent developments on the black hole information puzzle we refer the reader to \cite{raju2021BHIP, almheiri2020review}.

\begin{figure}[hptb]
		\centering
		\includegraphics{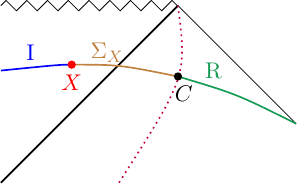}
		\caption{A schematic representation of the extremal surface $X$ (thick red dot) and the spacelike region $\Sigma_X$, denoted by the brown curve. The island region (I) is represented in blue, and the radiation region (green region) is denoted as R with $\pd$R = $C$. The thick black line is the horizon of the black hole, and the dotted purple line is an imaginary surface that defines the ``black hole region''. It continues all the way to past infinity and beyond this surface we have the ``radiation region''.}
		\label{fig:island_formula}
	\end{figure}

	One of the major breakthroughs towards finding a resolution to the information puzzle came in 2019 in the context of Jackiw-Teitelboim gravity \cite{teitelboim1983JT, jackiw1984JT} in AdS${}_2$ \cite{penington2019EWR, almheiri2019BHIP, almheiri2019Page, penington2019replicaWH, almheiri2019replicaWH, almheiri2019island}. The resolution highlighted that the entropy computation of Hawking -- which led to the information puzzle -- used the wrong notion of entropy. Instead a more careful analysis using the gravitational path integral to compute the fine-grained entropy \cite{ryu2006RT, ryu2006HEE, hubeny2007HRT, lewkowycz2013GGE, faulkner2013HEE, engelhardt2014QES} showed that new non-perturbative saddles in the gravitational path integral, called the ``replica wormholes'' \cite{almheiri2019replicaWH, penington2019replicaWH}, dominate over the ``Hawking saddle'' at late times and give a fine-grained entropy consistent with unitary time evolution \cite{page1994BHI, page2013HawkRad}. This computation implied that the von Neumann entropy of the radiation receives corrections from a spatially disjoint region called the ``island'' which contained the modes entangled with the Hawking radiation. When these additional corrections are accounted for, the radiation entropy does not grow with time and we have a behaviour consistent with unitarity. The formula for the gravitational fine-grained entropy is given by
	\begin{equation} \label{eq:island_formula}
		S = \underset{X}{\min} \left[ \underset{X}{\text{ext}} \left( \frac{\text{Area(X)}}{4G_N} + S_{\text{vN}}(\Sigma_X) \right) \right],
	\end{equation}
	where the surface $X$ is a codimension-2 surface, $\Sigma_X$ is a spacelike region bounded by $X$ and some cut-off surface. The LHS is the fine-grained entropy of the quantum state of the system under consideration (black hole or radiation), whereas the von Neumann entropy on the RHS is the entropy of the quantum state on $\Sigma_X$ in a semi-classical description. For instance, if $S$ were the entropy of radiation, the region $\Sigma_X$ would be the spacelike region bounded by $X$ and the boundary of the radiation region R (see Figure \ref{fig:island_formula}), or equivalently in a pure state, the complement of this spacelike region I $\cup$ R with $\pd$I = $X$. The surface which extremizes the fine-grained entropy is called the \textit{quantum extremal surface} (QES).

	While the Euclidean gravitational path integral computation for the unitary Page curve explains the phase transition in the entropy near Page time via the exchange of dominant saddle, a Lorentzian description for this transition is not completely well-understood. In \cite{pedraza2021Wald, hirano2023WaldIsland}, it was suggested that the generalized entropy in the island formula can be obtained as the Wald-like entropy \cite{wald1993entropy, iyer1994wald, jacobson1993JKM} in a Lorentzian description without making reference to an exterior non-gravitating bath region. The generalized entropy in the island formula is given as the Noether-Wald charge for the total gravitational action which incorporates the backreaction of the Hawking radiation in terms of a conformal anomaly term in the action via the Polyakov-Liouville action \cite{polyakov1981action, christensen1977anomaly}. The prescription was applied to Jackiw-Teitelboim gravity and the Russo-Susskind-Thorlacius model \cite{russo1992RST}, and it was shown that it gave a generalized entropy which reproduced the right Page curve already studied in these models via the island rule \cite{almheiri2019Page, penington2019EWR, hartman2020islands}.

	The original motivation behind the current work was to apply the Wald-like entropy prescription for islands in the 4D setup. However, in this paper, we only describe the Wald-like prescription for the generalized entropy of the quantum-corrected black hole solution in the dimensionally reduced Einstein-Hilbert gravity model \cite{buric1998Schwd2D, fabbri2005evaporation, djordjevic2022eternal, djordevic2025collapse, djordevic2025evaporating}. This theory admits both eternal and evaporating black hole solutions. In this article we will consider the eternal black hole and the quasi-stationary regime of the evaporating black hole. We show that the QES for the quantum-corrected eternal black hole is located at order $(c G_N)^2$ away from the backreacted horizon similar to the s-wave sector QES in the 4D Schwarzschild black hole in \cite{hashimoto2020Schwd}, which is also located at order $(c G_N)^2$ from the Schwarzschild radius of the classical black hole. However, it is interesting to note that when backreaction is considered the QES appears at the linear order away from the classical Schwarzschild radius.
	
	We also find the QES in the quasi-stationary regime of the evaporating black hole.\footnote{A more complete treatment of the dynamics of the gravitational collapse and the eventual evaporation of the black hole in the DREH model can be found in \cite{djordevic2025collapse, djordevic2025evaporating}.} In the quasi-stationary regime the black hole mass changes slowly and the solution we find in this approximation differs from the one obtained in \cite{djordevic2025collapse}. Because the Hawking evaporation timescale is parametrically much longer than the black hole's light-crossing time, the process can be described by treating the black hole as an equilibrium configuration at each instant \cite{hawking1975particle}. This approximation should be sufficient for semi-classical treatments of evaporation, but breaks down towards the final stages of black hole evaporation. In this approximation we find that, we have a family of QES that remains close to the event horizon at order $O(c G_N)$, unlike a single QES following a causal trajectory as is the case with the eternal black hole.

	We conclude this section by outlining the structure of the article: in Section \ref{sec:Wald_islands} we describe the Wald-like entropy proposal for the generalized entropy in the island formula. Section \ref{sec:DREH} summarizes the quantum-corrected eternal and evaporating black hole solutions in the DREH model, with explicit details of the computation in Appendix \ref{sec:bh_soln}. In Sections \ref{sec:eternal_islands} and \ref{sec:evaporating_islands}, we describe the computation of the generalized entropy for the island and no-island case in the eternal and the evaporating black hole, respectively. We also compute the location of the QES, the Page curves and the Page time in both these cases. We conclude in section \ref{sec:discussion} by describing some possible extensions to apply this formalism in four dimensions.

	\section{Brief review of Wald-like entropy for islands} \label{sec:Wald_islands}
	In a diffeomorphism-invariant theory of gravity, the conserved Noether currents $J^a$, associated with the diffeomorphism generators $\zeta^a$, can be expressed as \[ J^a = \nabla_b Q^{ab} \qq{with} Q^{ab} = - Q^{ba}, \] for some Noether potential $Q^{ab}$ \cite{wald1993entropy, iyer1994wald, jacobson1993JKM}. With a particular choice of the generator $\zeta^a$ we can define a Wald-like entropy as the Noether charge given by
	\begin{equation} \label{eq:Wald-like}
		S_{\text{Wald}} = 2\pi \int_X\ \dd^{D-2} x\, \epsilon_{ab}\, Q^{ab}[\zeta, \nabla \zeta, \cdots],
	\end{equation}
	where $X$ is a codimension-2 surface with binormal $\epsilon_{ab}$ such that $\epsilon_{ab} \epsilon^{ab} = -2$. This is a generalization of the Wald entropy for black holes, where in general, the Noether potential could depend on higher derivatives of the field $\zeta$. However, if $\zeta$ is a Killing field, say $\xi$, then the Noether potential depends only on $\xi^a, \nabla^b \xi^a$. For stationary black holes, $X$ is a spacelike cross-section of the horizon and $\xi^a$ corresponds to the horizon-generating Killing field. In fact, Wald's entropy is independent of the choice of cross-section $X$ and can be evaluated on the bifurcation surface, on which $\xi^a = 0, \nabla_{[a} \xi_{b]} = \kappa  \epsilon_{ab}$, where $\kappa$ is the surface gravity.

	To compute the generalized entropy of the island formula in \eqref{eq:island_formula} from a Wald-like entropy, we adopt the following prescription \cite{pedraza2021Wald, pedraza2021microcanonical, hirano2023WaldIsland}:
	\begin{equation} \label{eq:Sgen_Wald}
		S_{\text{gen}}(X, \Sigma_X) = 2\pi \int_X\ \dd^{D-2} x\, \epsilon_{ab}\, Q^{ab}[\xi^c, \nabla^c \xi^d] \bigg|_{\xi^c \to 0, \nabla^{[c}\xi^{d]} \to \epsilon^{cd}}\, ,
	\end{equation}
	where $X$ is the extremal surface, and we define
	 \begin{equation} \label{eq:vector_field}
		\xi^a \propto \epsilon^{ab}\, \nabla_b S_{gen}(X, \Sigma_X).  
	\end{equation}
	The normalization is fixed by demanding  $\nabla^{[a} \xi^{b]} = \epsilon^{ab}$ on $X$. Clearly, $\xi^a$ will be null and vanishing on $X$ by virtue of $X$ being an extremal surface. Moreover, $\xi^a = 0$ on $X$ and normal to it, this allows us to express $\nabla^{[a} \xi^{b]}$ as $\kappa \epsilon^{ab} + t^{[a} \xi^{b]}$ with $t^a$ being the tangent to $X$ and $\kappa$ is the analogue of the surface gravity in case of stationary black holes \cite{jacobson1993JKM}. On $X$, $\xi^a = 0$ allows us to pick the normalization of $\xi$ such that $\nabla^{[a} \xi^{b]} = \epsilon^{ab}$.

	We make some comments about this prescription before using it in the subsequent sections:
	\begin{enumerate}
		\item Compared to Wald's formalism where $X$ was a cross-section of the Killing horizon, in this case $X$ is not a cross-section of any horizon.
		\item $\xi$ was the horizon-Killing generator in Wald's formalism. However, in this case $\xi$ neither generates the extremal surface, nor is it Killing.
		\item It might seem that the definition of $S_{\text{gen}}$ is somewhat circular as we used $S_{\text{gen}}$ (in $\xi$) to define itself. However, to begin with, the prescription does not require the explicit expression for $S_{gen}$. The eventual existence of the extremal surface coming from an extremization of $S_{\text{gen}}$ as defined above is well-defined and self-consistent.
	\end{enumerate}

	\subsection{Wald-like entropy in dilaton-gravity theories} \label{sec:DG}
	Two-dimensional theories have been the playground for significant development towards a better understanding of the black hole information puzzle \cite{fiola1994GSL, callan1992CGHS, russo1992RST, penington2019EWR, penington2019replicaWH, almheiri2019BHIP, almheiri2019Page, almheiri2019replicaWH}. These theories are useful tools to infer some general features of the more difficult physics of the four-dimensional spacetime. It is often the case in spherically symmetric systems that the lowest angular momentum mode ($l  = 0$), or the s-wave sector dominates the physical effects. In particular, the truncation to the s-wave sector of the physical 4D black hole is well described by a 2D black hole in the resultant dimensionally reduced dilatonic gravity theory. A detailed account on dilaton gravity can be found in \cite{grumiller2002dilaton, fabbri2005evaporation}.

	We will be interested in applying the Wald-like prescription for generalized entropy to black hole solutions in 2D dilaton gravity obtained from the dimensional reduction of the Einstein-Hilbert action in four dimensions. The general form of the action of a 2D dilaton gravity theory we will be using is given by \cite{grumiller2002dilaton}
	\begin{equation} \label{eq:dilaton_gravity}
		I_{\text{DG}} = \frac{1}{16 \pi G_N} \int \dd^2 x\, \sqrt{-g}\, \left[ F(\phi) R + U(\phi) (\nabla \phi)^2 + V(\phi)  \right],
	\end{equation}
	where $\phi$ is the dilaton, and $(F(\phi), U(\phi), V(\phi))$ are arbitrary functions of the dilaton. Some examples include JT gravity, Mandal-Sengupta-Wadia/Callan-Giddings-Harvey-Strominger model \cite{mandal1991MSW, callan1992CGHS}, and RST model with following choices of these functions:
	\begin{equation} \label{eq:examples}
		(F(\phi), U(\phi), V(\phi)) = \begin{cases} (\phi, 0, 2 \phi) & \qq{JT gravity} \\ (e^{-2\phi}, 4e^{-2\phi}, 4 e^{-2\phi}) & \qq{MSW-CGHS model} \\ (e^{-2\phi} - \frac{c}{24} \phi, 4e^{-2\phi}, 4 e^{-2\phi}) & \qq{RST model} \end{cases}
	\end{equation}
	
	We will minimally couple the dilaton gravity to a 2D conformal field theory with central charge $c$, and quantum corrections will be incorporated via the Polyakov-Liouville action in terms of the auxiliary field called the \textit{conformalon} $\psi$, given by \cite{polyakov1981action}
	\begin{equation} \label{eq:PL_action}
		I_{\text{PL}} = - \frac{c}{24\pi} \int \dd^2 x\, \sqrt{-g}\, \left[ (\nabla \psi)^2 + \psi R \right],
	\end{equation}
	with the conformalon satisfying
	 \begin{equation} \label{eq:conformalon_eqn}
		2\, \Box\, \psi = R,
	\end{equation}
	such that the non-local Polyakov anomaly action can be cast in a local form as in \eqref{eq:PL_action}. So the total action can be written as
	\begin{equation} \label{eq:total_action}
		I = I_{\text{DG}} + I_{\text{m}} + I_{\text{PL}},
	\end{equation}
	with some minimally coupled 2D CFT as the matter sector having action $I_{\text{m}}$.

	The Noether potential in the Wald-like entropy is given as\footnote{Appendix C of \cite{pedraza2021Wald} contains a detailed discussion on the covariant phase space formalism for 2D dilaton gravity.}
	\begin{equation} \label{eq:Noether_potential_formula}
		Q^{ab} = -E^{abcd} \nabla_c \xi_d + 2 \xi_d \nabla_c E^{abcd},
	\end{equation}
	where $E^{abcd} = \pdv{L}{R_{abcd}}$, $L$ being the total Lagrangian. The only relevant terms in the Lagrangian for this computation are proportional to $F(\phi) R$ and $\psi R$. Varying these terms gives the Noether potential as
	\begin{equation} \label{eq:Noether_potential_model}
		Q^{ab} = -\left( \frac{F(\phi)}{16\pi G_N} - \frac{c}{24\pi} \psi \right) \nabla^{[a} \xi^{b]} + 2 \xi^{[a} \nabla^{b]} \left( \frac{F(\phi)}{16\pi G_N} - \frac{c}{24\pi} \psi \right).
	\end{equation}
	The Wald-like entropy from \eqref{eq:Sgen_Wald}, for the total action in \eqref{eq:total_action} is
	\begin{equation} \label{eq:Waldlike_Sgen}
		S_{\text{gen}}(X, \Sigma_X) = \left[ \frac{F(\phi)}{4G_N} - \frac{c}{6} \psi \right]_{X},
	\end{equation}
	where we evaluate the RHS at the location of the quantum extremal surface $X$. The precise form of $\phi$ and $\psi$ depends on the choice of the vacuum state, and are thus evaluated in the coordinates defining the corresponding vacuum. 

\begin{figure}[hptb]
		\centering
		\includegraphics{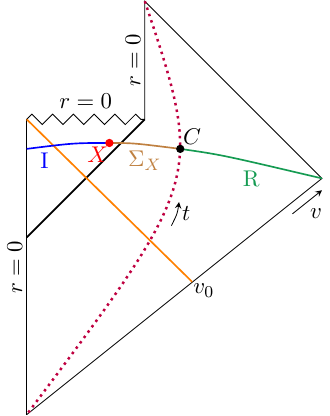}
		\caption{An imaginary timelike cut-off surface is shown as the purple dotted line. This surface separates the black hole region and the radiation region. After the Page time, a particular time-slice has three regions: the island region I, shown in blue, a disconnected region $\Sigma_X$ shown in brown, and the radiation region R in green. The boundary of the island region is the quantum extremal surface $X$. The boundary of the radiation region R is shown as the black dot $C$. Since we have a pure state, $S(\text{I} \cup \text{R}) = S(\Sigma_X)$.}
		\label{fig:evap_island}
	\end{figure}

	It should be emphasized that the above formula for $S_{\text{gen}}$ to determine the island applies to the case when $X$ is non-empty. The no-island extremum has no first-principle ``derivation'' from the Wald-like entropy, mainly because the Wald-like entropy does not distinguish between the geometric area contribution and the matter von Neumann entropy in the generalized entropy that enters the island formula in \eqref{eq:island_formula}. Hence, we cannot single-out the matter von Neumann entropy from the Noether-Wald charge in the absence of islands. However, when $X = \emptyset$, the time-dependent radiation entropy is actually captured by the conformalon contribution \cite{almheiri2014AdS2, pedraza2021Wald}
	\begin{equation} \label{eq:no_island}
		S_{\text{vN}}(\text{R}) = -\frac{c}{6} \psi\big|_{\pd\text{R}},
	\end{equation}
	where the RHS is evaluated at the boundary of the surface where the radiation is collected.\footnote{More precisely, the conformalon field is evaluated at the ``image point'' of $\pd$R = $C$ shown in Figures \ref{fig:eternal_island} and \ref{fig:evap_island}. In the case of the eternal black hole, the point of evaluation will be $C'$ which is the boundary of the radiation region on the left exterior. While in the evaporating black hole case it will be the image point of $C$ obtained by running the incoming null ray at $C$ back through the $r = 0$ surface, with reflecting boundary conditions.} For the eternal black hole in JT gravity \cite{pedraza2021Wald, hirano2023WaldIsland} and the evaporating black hole in RST model \cite{hirano2023WaldIsland}, it was shown that this prescription correctly gives the generalized entropy of the island formula and that the corresponding Page curve agrees with earlier results in the literature in \cite{almheiri2019Page, penington2019replicaWH, almheiri2019island, hartman2020islands}.
	
	The expression in \eqref{eq:no_island} is not a mere coincidence: the conformalon captures the information of the state and the data of the conformal transformation relating the physical coordinates to the coordinates in which the state looks like the vacuum. Moreover, the Polyakov-Liouville action is, by construction, the generating functional of exactly the same conformal anomaly term encoding the entanglement entropy of the radiation interval. In other words, the conformalon is the local field-theoretic realization of the same conformal anomaly data that the replica trick computes, and thus $-\frac{c}{6} \psi$ reproduces the Cardy–Calabrese formula \cite{calabrese2009CFT} for the entanglement entropy of the interval whose endpoints are set by the boundary of the radiation region.\footnote{The RHS of \eqref{eq:no_island} actually computes the von Neumann entropy of the complementary region $\Sigma_X$, on the same time-slice on which the pure state is defined, and we have $S(\text{I} \cup \text{R}) = S(\Sigma_X)$ with islands, or $S(\text{R}) = S(\Sigma_X)$ without islands.}

	\section{Dimensionally reduced Einstein-Hilbert gravity} \label{sec:DREH}
	In this section, we will describe the dimensionally reduced gravity model for the 4D Einstein-Hilbert action \cite{buric1998Schwd2D, djordjevic2022eternal, djordevic2025collapse, djordevic2025evaporating}, and describe black hole solution for the 1-loop quantum-corrected action when the Polyakov-Liouville term is included. To simplify the arguments, we will assume that the massless matter is the dominant contribution to the Hawking radiation. We model our matter sector with $N$ number of minimally-coupled massless scalar fields $\vec{\chi} = (\chi_1, \ldots, \chi_N)$, and take the central charge $c = N$ to be large. The arguments are, however, more general and should be applicable to any minimally-coupled CFT with large central charge at the semi-classical level.

	The 4D Einstein-Hilbert action is given by
	\begin{equation} \label{eq:EH_action}
		I_{\text{EH}} = \frac{1}{16\pi G_N^{(4)}} \int \dd^4 x\, \sqrt{-\det g^{(4)}} R^{(4)},
	\end{equation}
	where, $G_N^{(4)}$ is the 4D Newton's constant, $g^{(4)}_{\mu \nu}$ is the 4D metric and the corresponding Ricci scalar is given by $R^{(4)}$. We will consider a spherically symmetric metric ansatz of the form
	\begin{equation} \label{eq:4D_metric}
		\dd s^2 = g^{(4)}_{\mu \nu} \dd x^{\mu} \dd x^{\nu} = g_{ab}(x^c) \dd x^{a} \dd x^{b} + \lambda^{-2} e^{-2\phi(x^c)} \dd \Omega^2_2,
	\end{equation}
	where, $\dd \Omega_2^2$ is the standard metric on the unit sphere, and the dilaton field $\phi$ is related to the radial coordinate via
	\begin{equation} \label{eq:radial_dilaton}
		r = \frac{1}{\lambda} e^{-\phi},
	\end{equation}
	with $\lambda^2$ being a constant that will play the role of ``cosmological constant'' in the 2D dilaton gravity theory. The dimension of $\lambda$ is that of inverse length.

	We can then reduce the 4D Ricci scalar $R^{(4)}$, in terms of the 2D Ricci scalar $R$ and the dilaton field as
	 \begin{equation} \label{eq:Ricci_reduction}
		 R^{(4)} = R - 6 (\nabla \phi)^2 + 4 \Box{\phi} + 2 \lambda^2 e^{2 \phi}.
	\end{equation}
	Then, up to a surface term the dimensionally reduced Einstein-Hilbert action is
	\begin{equation} \label{eq:DREH_action}
		I_{\text{DREH}} = \frac{1}{16\pi G_N} \int \dd^2 x\, \sqrt{-\det g} \left[ e^{-2\phi} (R + 2 (\nabla \phi)^2) + 2 \lambda^2 \right],
	\end{equation}
	where we have introduced the 2D Newton's constant $G_N$, as $4\pi G_N = \lambda^2 G_N^{(4)}$. With the matter sector taken into account the total classical action is
	\begin{equation} \label{eq:classical_action}
		\begin{split}
			I_{\text{cl}} &= I_{\text{DREH}} + I_{\text{m}} \\ &= \frac{1}{16\pi G_N} \int \dd^2 x\, \sqrt{-\det g} \left[ e^{-2\phi} (R + 2 (\nabla \phi)^2) + 2 \lambda^2 \right] - \frac{1}{2} \int \dd^2 x\, \sqrt{-\det g}\, g^{ab}\, \nabla_a \vec{\chi}\, \cdot \nabla_b \vec{\chi}.
		\end{split}
	\end{equation}
	The classical field equations obtained from this action are
	\begin{align}
		g_{ab} &: \quad e^{-2\phi} \left[ 2\nabla_a \nabla_b \phi - 2\nabla_a \phi \nabla_b \phi + g_{ab} (3 (\nabla \phi)^2 - 2\Box \phi - \lambda^2 e^{2\phi}) \right] = 8\pi G_N \T_{ab,\, \text{cl}}, \label{eq:metric_eom_cl} \\
		\phi &: \quad \Box \phi - (\nabla \phi)^2 = -\frac{R}{2}, \label{eq:dilaton_eom_cl} \\
		\vec{\chi} &: \quad \Box \vec{\chi} = 0 \label{eq:matter_eom_cl},
	\end{align}
	where the classical stress tensor is given as
	\begin{equation} \label{eq:classical_EM}
		\T_{ab,\, \text{cl}} = - \frac{2}{\sqrt{-\det g}} \fdv{I_{\text{m}}}{g^{ab}} = \sum_{i = 1}^N \nabla_a \chi_i \nabla_b \chi_i - \frac{g_{ab}}{2} (\nabla \vec{\chi})^2.
	\end{equation}
	
	To study the quantum corrections to this setup we introduce the Polyakov-Liouville term in the action, so that the total action at the 1-loop level is given by
	\begin{equation} \label{eq:1loop_action}
		I = I_{\text{cl}} + I_{\text{PL}},
	\end{equation}
	with $I_{\text{PL}}$ as in \eqref{eq:PL_action}. Varying $I_{\text{}PL}$ with respect to the metric gives the quantum correction to the stress tensor for scalar fields
	\begin{equation} \label{eq:1loop_EM}
		\ev{\T_{ab}} = \frac{c}{12\pi} \left[ -\nabla_a \nabla_b \psi + \nabla_a \psi \nabla_b \psi + g_{ab} \left( \Box \psi - \frac{1}{2} (\nabla\psi)^2 \right) \right],
	\end{equation}
	where in evaluating the expectation value we also need to specify the quantum state of the matter, whose information is encoded in the conformalon solution. The total stress tensor at the 1-loop level is then given by
	\begin{equation} \label{eq:1loop_EM_total}
		\T_{ab} = \T_{ab,\, \text{cl}} + \ev{\T_{ab}}.
	\end{equation}
	The second term will give an additional contribution to the metric field equations in \eqref{eq:metric_eom_cl}
	\begin{equation} \label{eq:metric_eom_quantum}
		\begin{split}
			&e^{-2\phi} \left[ 2\nabla_a \nabla_b \phi - 2\nabla_a \phi \nabla_b \phi + g_{ab} (3 (\nabla \phi)^2 - 2\Box \phi - \lambda^2 e^{2\phi}) \right] \\ &= 8\pi G_N \sum_{i = 1}^N \nabla_a \chi_i \nabla_b \chi_i - \frac{g_{ab}}{2} (\nabla \vec{\chi})^2 +  \frac{2cG_N}{3} \left[ -\nabla_a \nabla_b \psi + \nabla_a \psi \nabla_b \psi + g_{ab} \left( \Box \psi - \frac{1}{2} (\nabla\psi)^2 \right) \right].
		\end{split}
	\end{equation}
	We will treat the dilaton as a purely classical field, hence the dilaton equation of motion remains unaffected by including the 1-loop Polyakov-Liouville term in the action. It will be interesting to remove this restriction, and in the 4D case there are indeed dilaton-conformalon couplings arising from 1-loop correction \cite{mukhanov1994HawkRad, fabbri2005evaporation}. Since the model in \eqref{eq:1loop_action} is not exactly solvable we will find a linearized quantum backreacted solution about the classical vacuum \cite{buric1998Schwd2D, frolov1996oneloppchargedBH, fabbri2005evaporation}, where the backreaction is controlled by the natural semi-classical parameter \cite{hartle1981semiclassics, flanagan1996semiclassics}
	\begin{equation} \label{eq:expansion_parameter}
		\varepsilon = \frac{2cG_N}{3} \ll (\lambda r_0)^2,
	\end{equation}
	with $c \gg 1$ and $G_N \to 0$, and $r_0$ being the classical Schwarzschild radius of the black hole. This model admits both eternal and evaporating black hole solutions. We give a self-contained description of the eternal and the quasi-stationary evaporating black hole solutions in Appendix \ref{sec:bh_soln}, and below we summarize the key features of the solutions.

	\subsection{Eternal black hole} \label{sec:eternal}
	\begin{figure}[hptb]
		\centering
		\includegraphics{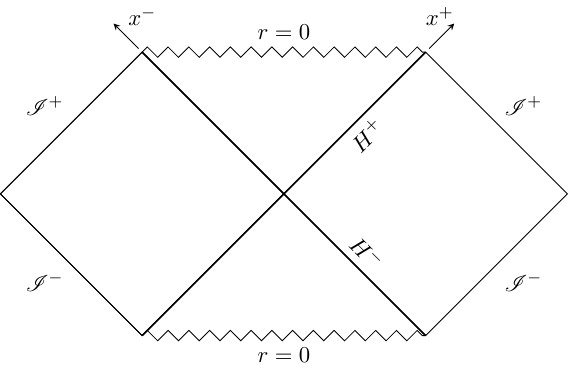}
		\caption{Penrose diagram of the eternal black hole spacetime in the DREH model. The thick black lines indicate the horizons located at $x^+ = 0$ or $x^- = 0$.}
		\label{fig:eternal_bh}
	\end{figure}
	In this subsection we will summarize the quantum-corrected eternal black hole solution which corresponds to Hartle-Hawking state \cite{hartle1976hawking, israel1976TFD}. For a distant observer, the black hole appears to be in a thermal equilibrium with the Hawking radiation in the Hartle-Hawking state. The backreacted geometry takes the following form
	\begin{equation} \label{eq:1loop_eternal_bh}
		\dd s^2 = -f(r) e^{2\varepsilon h(r)} \dd t^2 + \frac{\dd r^2}{f(r)} = -e^{2\rho(x)} \dd x^+ \dd x^-,
	\end{equation}
	with \[ f(r) = 1 - \frac{r_0}{r} + \varepsilon \frac{m(r)}{r}, \quad r = \lambda^{-1} e^{-\phi}, \] and $x^\pm = \pm \kappa^{-1} e^{\pm (t\, \pm\, r_*)}$ being the Kruskal-Szekeres coordinates. The different functions appearing in the metric coefficients are given by
	\begin{equation} \label{eq:metric_coeff_eternal}
		\begin{split}
			m(r) &= \frac{1}{8\lambda^2 r_0} \left( -\frac{7}{2} \frac{r_0^2}{r^2} + \frac{r_0}{r} - \log \frac{r}{L} - \frac{r}{r_0} + \frac{L}{r_0} \right), \\
			h(r) &= \frac{1}{8\lambda^2 r_0^2} \left( -\frac{3}{2} \frac{r_0^2}{r^2} - 2 \frac{r_0}{r} + \log \frac{r}{L} \right), \\
			\rho(x) &= \frac{1}{2} \log f + \varepsilon h - \kappa r_*, \\
			\dd r_* &= \frac{\dd r}{f(r) e^{\varepsilon h(r)}}.
		\end{split}
	\end{equation}
	where $\kappa$ is the surface gravity, $r_*$ is the tortoise coordinate, and we use $r_0 = \frac{8\pi G_N M}{\lambda^2}$ to denote the Schwarzschild radius of the classical black hole solution, while $r_h = r_0 - \varepsilon m(r_0)$ is the location of the horizon in the quantum-corrected black hole solution and is given by
	\begin{equation} \label{eq:eternal_corrected_horizon}
		r_h = r_0 + \frac{\varepsilon}{8\lambda^2 r_0} \left( \frac{7}{2} - \frac{L}{r_0} + \log \frac{r_0}{L} \right).
	\end{equation}
	We also introduced a large-distance length scale $L$ such that \[ \lim_{L \to \infty} \lim_{r \to L} m(r) = 0 \qq{and} \lim_{L \to \infty} \lim_{r \to L} h(r) = 0. \]
	
	The full conformalon field in the Hartle-Hawking state is given by
	\begin{equation} \label{eq:conformalon_eternal}
		\psi(x) = -\rho(x) + \psi_+(x^+) + \psi_-(x^-),
	\end{equation}
	and we can solve for the homogeneous part $\psi_\pm(\pm)$, of the conformalon from \eqref{eq:t_pm_defn} and \eqref{eq:HH_vac} such that
	\begin{equation} \label{eq:conformalon_homo_sc_eternal}
		\psi_\pm(x^\pm) = - \log(x^\pm - c^\pm) - d^\pm,
	\end{equation}
	where $c^\pm, d^\pm$ are integration constants. We will fix $c^\pm$ by requiring that they represent the coordinates of a point on the cut-off surface, which defined the black hole and the radiation regions as shown in Figure \ref{fig:eternal_island}. In particular, the cut-off surface itself is defined by the product $c^+ c^- = constant$, such that the metric remains a constant on the cut-off surface. The point $(c^+, c^-)$ will then be the intersection of the cut-off surface and a time-slice $t = t_c$ \cite{pedraza2021Wald, hirano2023WaldIsland}. This will also require us to regularize the behaviour of the $\psi_\pm(x^\pm)$ as $x^\pm \to c^\pm$, this can be done by introducing a short-distance length scale $\eta$, which regulates this limit \cite{pedraza2021Wald}. This length scale will not be directly relevant to our results, but it is nice to see that this is an artefact of the same UV divergence that features in the von Neumann entropy in quantum field theories. Further, by requiring local Lorentz invariance of the scalar $\psi$, we can fix the sum $d^+ + d^-$ as $\rho(r_c) \equiv \rho_c$. The conformalon solution in the Hartle-Hawking state is then given by
	\begin{equation} \label{eq:conformalon_HH}
		\psi(x) = -\frac{1}{2} \log [e^{2\rho(x)} e^{2\rho_c} (x^+ - c^+)^2(x^- - c^-)^2 \eta^{-4}].
	\end{equation}

	\subsection{Evaporating black hole} \label{sec:evaporating}
	\begin{figure}[hptb]
		\centering
		\includegraphics{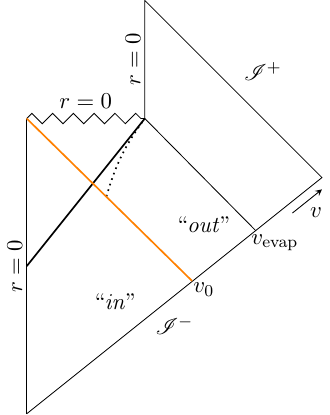}
		\caption{Schematic Penrose diagram for black hole formation from collapse of a null shell of mass $M_0$ at $v = v_0$ (thick orange line), and its eventual evaporation due to Hawking radiation in the DREH model. The event horizon is shown by the thick black line, while the apparent horizon is the dotted line. The region before the surface of formation $v = v_0$, is Minkowski spacetime (``in'' region), and the region $v_0 < v < v_{\text{evap}} \sim O(M_0^3)$ is the black hole spacetime (``out'' region). The region $v > v_{\text{evap}}$ is again flat spacetime, with the surface $v = v_{\text{evap}}$ indicating where the complete evaporation of the black hole has taken place. The (linearized) semi-classical analysis in this work is not sufficient to capture the physics close to the endpoint of the evaporation, although it is a good approximation in the region $v < v_{\text{evap}}$.}
		\label{fig:evaporating_bh}
	\end{figure}
	The full process of dynamical gravitational collapse of matter forming a black hole and its eventual evaporation is described by the ``$in$'' state.\footnote{For a detailed discussion on the different choices of vacuum states we refer the reader to \cite{birrell1984, fabbri2005evaporation}.} The $\ket{in}$ state corresponds to the Minkowski vacuum on past null infinity, and to the thermal radiation at late times. The latter is also described by the Unruh vacuum \cite{unruh1976effect} defined with respect to the positive frequency modes of $(U, v)$, where we denote the Eddington-Finkelstein coordinates as $(u, v)$, and the Kruskal coordinate as $U$. Since we will not be concerned with the specific details of the collapse phase, to describe the quantum-correction we will work in the late time quasi-stationary regime where the two states agree on the metric backreaction \cite{fabbri2005evaporation}.

	We consider the quantum backreaction of the matter on the black hole formed by the collapse of a null shell of matter. The backreacted geometry takes the following form
	\begin{equation} \label{eq:1loop_evap_bh}
		\dd s^2 = -f(r, v) e^{2\varepsilon h(r, v)} \dd v^2 + 2 \dd v \dd r = -e^{2\rho(u, v)} \dd u \dd v,
	\end{equation}
	such that $v > v_0$ and \[ f(r, v) = 1 - \frac{r_0}{r} + \varepsilon \frac{m(r, v)}{r}, \quad r = \lambda^{-1} e^{-\phi}. \] The different functions appearing in the metric coefficients are given by 
	\begin{equation} \label{eq:metric_coeff_evap}
		\begin{split}
			m(r, v) &= \frac{1}{8\lambda^2 r_0} \left( -\frac{7}{2} \frac{r_0^2}{r^2} + \frac{r_0}{r} - \log \frac{r}{L} - \frac{r}{r_0} + \frac{L}{r_0} + \frac{v - v_0}{2r_0} \right), \\
			h(r, v) &= \frac{1}{8\lambda^2 r_0^2} \left( -\frac{3}{2} \frac{r_0^2}{r^2} - 2 \frac{r_0}{r} + \log \frac{r}{L} \right), \\
			\rho(u, v) &= \frac{1}{2} \log f(r(u, v), v) + \varepsilon h(r(u, v), v).
		\end{split}
	\end{equation}
	where $r_0$ and $\kappa_0$ are respectively the classical Schwarzschild radius and the surface gravity of the black hole with
	\begin{equation} \label{eq:evap_radius_surface_gravity}
		r_0 = \frac{8\pi G_N M_0}{\lambda^2} \qq{and} \kappa_0 = \frac{1}{2r_0}.
	\end{equation}
	The backreacted apparent and event horizons are located at
	\begin{equation} \label{eq:event_apparent_hors}
		\begin{split}
			r_h^{\text{AH}}(v) &= r_0 + \frac{\varepsilon}{8\lambda^2 r_0} \left( \frac{7}{2} - \frac{L}{r_0} + \log \frac{r_0}{L} - \frac{(v - v_0)}{2r_0} \right), \\
			r_h(v) &= r_h^{\text{AH}}(v) - \frac{\varepsilon}{8\lambda^2 r_0}\, .
		\end{split}
	\end{equation}
	
	The full conformalon field in the Unruh state is given by
	\begin{equation} \label{eq:conformalon_evap}
		\psi(u, v) = -\rho(u, v) + \psi_1(v) + \psi_2(u),
	\end{equation}
	The homogeneous parts $\psi_1(v), \psi_2(u)$ can be solved from \eqref{eq:t_v} and \eqref{eq:t_u} as
	\begin{equation} \label{eq:conformalon_homo_sc_evap}
		\begin{split}
			\psi_1(v) &= -\log(v - v_1) - v_2 \\
			\psi_2(u) &= -\log(\sinh(\frac{\kappa_0}{2} (u - u_1))) - u_2.
		\end{split}
	\end{equation}
	where $v_1, v_2, u_1, u_2$ are integration constants. Similar to the eternal black hole case, we will fix $(u_1, v_1) \equiv (u_c, v_c)$ as the coordinates of a point on the cut-off surface, where the point itself is located at the intersection of the cut-off surface and a time-slice $t(u_1, v_1) \equiv t_c$. Also, from local Lorentz invariance of the scalar $\psi$ we require $u_2 + v_2 = \rho(u_c, v_c) + \log \frac{2}{\kappa_0}$. The conformalon solution in the Unruh state is then given by
	\begin{equation} \label{eq:conformalon_uv}
		\psi(u, v) = -\frac{1}{2} \log\! \left[ \frac{4}{\kappa_0^2} e^{2\rho(u, v)} e^{2\rho(u_c, v_c)} (v - v_c)^2 \sinh^2 \left( \frac{\kappa_0}{2} (u - u_c) \right) \right],
	\end{equation}
	In the $(U, v)$ coordinates, which is the more natural set of coordinates to express the conformalon in the Unruh state we get
	\begin{equation} \label{eq:conformalon_U}
		\psi(U, v) = -\frac{1}{2} \log\! \left[ e^{2\rho(U, v)} e^{2\rho_c} (v - v_c)^2 (U - U_c)^2 \eta^{-4} \right],
	\end{equation}
	where $\rho_c \equiv \rho(U_c, v_c) = \rho(u_c, v_c) + \frac{1}{2} \kappa_0 u_c$, and we have introduced the short distance regulator (defined in the local Lorentz frame at $C$) $\eta$ similar to the Hartle-Hawking case in \eqref{eq:conformalon_HH}.

	Finally we should account for the presence of the surface $r = 0$, on which we consider reflecting boundary conditions. Because of the reflecting boundary conditions, there exists correlations between the left-movers and the right-movers, and $\psi_1(v)$, $\psi_2(u)$ are not genuinely independent data on the physical Hilbert space. We can establish a relation between the two solutions by looking at the map that relates an outgoing ray at retarded time $u$ traced back through the collapsing shell to an ingoing ray at some advanced time $v$ that bounced off $r = 0$.

	\section{Island and Page curve for eternal black hole in DREH model} \label{sec:eternal_islands}
	In this section, we will describe the island and Page curve for the eternal black hole in the DREH model. We first obtain the generalized entropy from the Wald-like entropy formula in \eqref{eq:Waldlike_Sgen} and extremize it to obtain the location of the QES. Then we describe the no-island case and show that we get the expected Page curve \cite{djordjevic2022eternal}.

	\subsection{Generalized entropy and location of island} \label{sec:eternal_Sgen_island}
	\begin{figure}[hptb]
		\centering
		\includegraphics{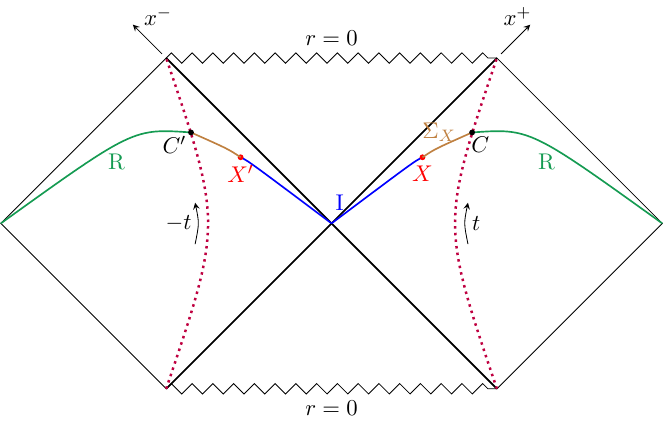}
		\caption{An imaginary timelike cut-off surface is shown as the purple dotted line on either exterior, and we take the curves to be symmetrically placed. This surface defines the ``black hole region'' (inside this surface) and the ``radiation region'' (outside this surface). A particular time-slice (beyond the Page time) has been shown which has three components: island region I (blue), the union of disconnected regions, denoted as $\Sigma_X$ (brown), and the radiation region R (green). The boundary of the island region is the quantum extremal surface $X$ on the right exterior with coordinates $(a^+, a^-)$ in the Kruskal-Szekeres chart or $(t_a, a)$ in the Schwarzschild chart, and $X'$ on the left exterior with coordinates $(a^+_L, a^-_L) = (a^-, a^+)$ or $(-t_a + i \frac{\pi}{\kappa}, a_*)$. The boundary of the radiation region R is shown as the black dot $C$. Since we have a pure state, $S(\text{I} \cup \text{R}) = S(\Sigma_X)$.}
		\label{fig:eternal_island}
	\end{figure}
	 From the Wald-like entropy relation in \eqref{eq:Waldlike_Sgen} we determine the generalized entropy in the Hartle-Hawking state to be
	 \begin{equation} \label{eq:HH_Sgen}
		 S_{\text{gen}}^{\text{HH}}(t_a, a_*; t_c, c_*) = 2 \times \left[ \frac{\lambda^2 r^2}{4 G_N} + \frac{c}{12} \log [e^{2\rho(r)} e^{2\rho_c} (x^+ - c^+)^2(x^- - c^-)^2 \eta^{-4}] \right]_{x^\pm = a^\pm},
	 \end{equation}
	 where the additional factor of $2$ is to account for the fact that we have a two-sided black hole with identical configuration for the two exteriors. We think of $r, x^\pm$ as functions of $t, r_*$, and the coordinates $a^\pm = \pm \kappa^{-1} e^{\pm \kappa (t_a \pm a_*)}$ give the location of the quantum extremal surface $X = \pd$I,\footnote{More precisely, $(a^+, a^-)$ are the coordinates of the right boundary of the island region I. The left boundary coordinates are given by $a^\pm_L = a^\mp$. Alternatively, in the Schwarzschild chart covering the left exterior the island boundary is located at $(t_L, r_L) = (-t_a + i \frac{\pi}{\kappa}, a)$.} with I being the island region, determined via
 	\begin{equation} \label{eq:extremization}
		\pd_t S_{\text{gen}}^{\text{HH}}(t, r_*) \bigg|_{a^\pm} = 0 \qq{and} \pd_{r_*} S_{\text{gen}}^{\text{HH}}(t, r_*) \bigg|_{a^\pm} = 0.
 	\end{equation}

	The extremization with respect to $t$ immediately gives
	 \begin{equation} \label{eq:eternal_t_extremization}
		t|_{a^\pm} \equiv t_a = t_c,
	\end{equation}
	so the extremal surface $X$ lies on the same time-slice as the point $(c^+, c^-)$ on the cut-off surface. Whereas the extremization with respect to $r_*$ results in the following relation
	 \begin{equation} \label{eq:eternal_r_extremization_relation}
		 \left( r + \frac{\varepsilon}{2\lambda^2} \dv{\rho(r)}{r} \right) f(r) e^{-\varepsilon h(r)} \bigg|_{r=a} = \frac{1}{\lambda^2} \frac{\varepsilon \kappa}{e^{\kappa(c_* - a_*)} - 1}
	\end{equation}
	To obtain the location $a$, we will assume that the QES will be located close to the backreacted horizon at $r = r_h$ \cite{penington2019EWR, almheiri2019island}. Then
	 \begin{equation} \label{eq:eternal_X_approx}
		 a = r_h + \delta, \qq{with} \delta \ll r_h.
	\end{equation}
	Then the only non-zero term on LHS of \eqref{eq:eternal_r_extremization_relation} will be
	 \begin{equation} \label{eq:eternal_r_extremization_relation_LHS}
		\delta \times \left( r_h + \frac{\varepsilon}{2\lambda^2} \rho'(r_h) \right) f'(r_h) e^{-\varepsilon h(r_h)} = \delta \times 2\kappa \left( r_h + \frac{\varepsilon}{2\lambda^2} \rho'(r_h) \right),
	\end{equation}
	where $' \equiv \dv{r}$. Similarly using \[ e^{2\kappa a_*} = \frac{\delta}{r_h} e^{1 + \varepsilon g(r_h)}\ \text{with}\ g(r) = \frac{3r}{8\lambda^2 r_h^3}, \] we can show that the RHS is given by
	\begin{equation} \label{eq:eternal_r_extremization_relation_RHS}
		\frac{\varepsilon \kappa}{\lambda^2} \left[ \sqrt{\frac{\delta}{r_h}} e^{\frac{1}{2} (1 + \varepsilon g(r_h)) - \kappa c_*} + \frac{\delta}{r_h} e^{1 + \varepsilon g(r_h) - 2\kappa c_*} \right]
	\end{equation}
	Thus solving for $\delta$ we get
	\begin{equation} \label{eq:eternal_delta}
		\delta = \frac{\varepsilon^2}{4\lambda^4 r_h^3} e^{1 - 2\kappa c_*}
	\end{equation}
	So the QES is located at
	\begin{equation} \label{eq:eternal_QES}
		t_a = t_c,\ a = r_h + \frac{(cG_N)^2}{9\lambda^4 r_h^3} e^{1 - 2\kappa c_*}.
	\end{equation}
	Note that we first extremized the Hartle-Hawking generalized entropy over time $t$ to get $t_a = t_c$, and used it in the formula for the Hartle-Hawking generalized entropy in \eqref{eq:HH_Sgen}, making it time-independent before extremizing over $r_*$ to get $a$. Instead we could have extremized the time-dependent Hartle-Hawking generalized entropy with respect to $r_*$ and still obtained the same extremal conditions as in \eqref{eq:eternal_QES}. 
	
	The QES is located away from the backreacted horizon and outside it at order $(cG_N)^2$, but at linear order from the original Schwarzschild radius $r_0$. This is interesting because in \cite{hashimoto2020Schwd} it was shown that, without the backreaction effects, the extremal surface is located at order $(cG_N)^2$ from the location of the horizon in the s-wave approximation. However, unlike \cite{hashimoto2020Schwd}, we also consider the backreaction of the matter in the s-wave sector, but the quantum corrections do not change the order at which the islands are located away from the (quantum-corrected) horizon in the backreacted geometry.	

	From \eqref{eq:HH_Sgen}, the fine-grained entropy in the Hartle-Hawking state to the leading order in $\varepsilon$ is given by
	\begin{equation} \label{eq:Sgen_HH_corrected}
		S_{\text{island}}^{\text{HH}} \equiv S_{\text{gen}}^{\text{HH}}(t_a, a_*; t_c, c_*) = 2 \times \left[ \frac{\lambda^2 r_h^2}{4 G_N} + \frac{c}{12} \log(e^{2\rho_H} e^{2\rho_c} \frac{e^{4 \kappa c_*}}{\kappa^4 \eta^4}) \right] \equiv 2S_{\text{bh}},
	\end{equation}
	where $\rho_H \equiv \rho(r_h)$, and $S_{\text{bh}}$ represents the entropy of the black hole region, which receives two contributions, namely, the Bekenstein-Hawking horizon entropy and the matter entropy within the cut-off surface. In particular, the Hartle-Hawking generalized entropy is independent of time.
	
	\subsection{No-island case} \label{sec:eternal_no_island}
	\begin{figure}[hptb]
		\centering
		\includegraphics{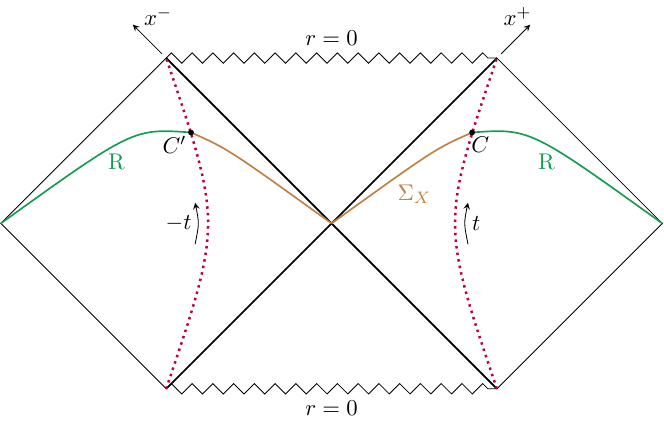}
		\caption{In the case of no-island, the entropy of the radiation $S(\text{R})$ is same as the entropy of the complementary region shown in brown and is encoded in the conformalon solution evaluated at the boundary of the radiation region $C'$ with coordinates $(-t_c + i \frac{\pi}{\kappa}, c_*)$, where $(t_c, c_*)$ are the coordinates of the other boundary $C$. Before the Page time the radiation entropy is time-dependent with respect to the time along the imaginary cut-off surface.}
		\label{fig:eternal_noisland}
	\end{figure}

	\noindent Recall that in the absence of island, the radiation entropy is given by the conformalon solution in the corresponding state as in \eqref{eq:no_island}. In the Hartle-Hawking state we obtained the conformalon solution in \eqref{eq:conformalon_HH} from which we infer the radiation entropy as
	\begin{equation} \label{eq:rad_entropy_no_island_eternal}
		\begin{split}
			S_{\text{no-island}}^{\text{HH}}(t_c, c_*) &= \frac{c}{12} \log [e^{2\rho(x)} e^{2\rho_c} (x^+ - c^+)^2(x^- - c^-)^2 \eta^{-4}]_{x^\pm = -c^\mp + i \frac{\pi}{\kappa}} \\
		   &= \frac{c}{12} \log(4 e^{4\rho_c} e^{4\kappa c_*} \frac{\cosh^4(\kappa t_c)}{\kappa^4 \eta^4}) \xrightarrow{\kappa t_c \gg 1} \frac{c}{3} \kappa t_c.
		\end{split}
	\end{equation}
	 We see that not including the islands give an unbounded entropy growth for the radiation at late times.

	\subsection{Page time and Page curve} \label{sec:eternal_Page}
	The fine-grained entropy of the radiation in the Hartle-Hawking state will obey a Page curve which is given by
	\begin{equation} \label{eq:Page_curve}
		\min\left\{ S_{\text{no-island}}^{\text{HH}}(t), S_{\text{island}}^{\text{HH}} \right\}.
	\end{equation}
	The difference of the island and no-island entropies is approximately
	\begin{equation} \label{eq:HH_entropy_diff}
		\frac{\lambda^2 r_0^2}{2G_N} - \frac{c}{3} \log \cosh(\kappa t) + O(\varepsilon).
	\end{equation}
	For small $\kappa t$, $\log \cosh(\kappa t) \approx (\kappa t)^2 $, and the difference is positive. So, initially, the no-island entropy is smaller. However, beyond the Page time $t_{\text{Page}}$, the no-island entropy will take over the island entropy. This crossing over takes place at the Page time which is approximately given by
	\begin{equation} \label{eq:Page_time_eternal}
		t_{\text{Page}} \approx \frac{1536 \pi^3 G_N^2 M^3}{c \lambda^4} + O(\varepsilon).
	\end{equation}

	\begin{figure}[hptb]
		\centering
		\includegraphics{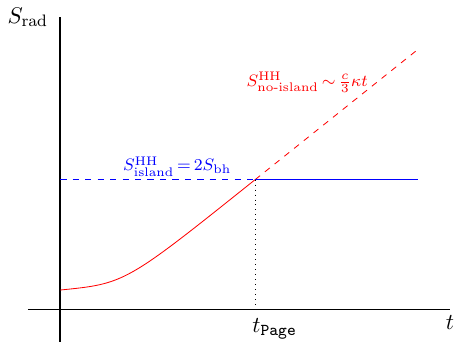}
		\caption{The Page curve for the radiation in the Hartle-Hawking state is indicated by the solid red and blue curves. For $\kappa t$ very small the behaviour of the fine-grained entropy is approximately quadratic in $t$, whereas at late times it becomes linear in $t$ and grows unboundedly. After Page time $t_{\text{Page}} \sim \frac{S_{\text{bh}}}{c \kappa}$, the fine-grained entropy of the radiation is given by the island contribution which is a constant.}
		\label{fig:eternal_Page}
	\end{figure}
	
	\section{Island and Page curve for evaporating black hole in DREH model} \label{sec:evaporating_islands}
	In this section, we will describe the island and Page curve for the quasi-stationary evaporating black hole in the DREH model. We will first obtain the generalized entropy from the Wald-like entropy formula in \eqref{eq:Waldlike_Sgen} and extremize it to obtain the location of the QES. Then we describe the no-island case and the Page curve for the evaporating black hole.

	\subsection{Generalized entropy and location of island} \label{sec:evap_Sgen_island}

	From the Wald-like entropy relation in \eqref{eq:Waldlike_Sgen} we determine the generalized entropy in the Unruh state to be
	\begin{equation} \label{eq:Unruh_Sgen}
	    S_{\text{gen}}^{\text{U}}(U_a, v_a; U_c, v_c) = \left[ \frac{\lambda^2 r^2}{4 G_N} + \frac{c}{12} \log\! \left[ e^{2\rho(U, v)} e^{2\rho_c} (v - v_c)^2 (U - U_c)^2 \eta^{-4} \right] \right]_{(U_a, v_a)},
	\end{equation}
	where in the first term $r$ is understood to be a function of $(U, v)$.

	In order to extremize $S_{\text{gen}}^{\text{U}}$ to find the location of the QES we will make some two simplifying assumptions motivated from \cite{penington2019EWR}:
	\begin{enumerate}
		\item We expect the island to lie close to the horizon, i.e. the radial location of the QES should be at $a = r_h(v_a) + \delta$, for $\frac{\delta}{r_h(v_a)} \ll 1$.
		\item Since we will be concerned with the region close to the horizon we can approximate the radial coordinate as
	   	\begin{equation} \label{eq:radial_approx}
	   		r(U, v) \simeq r_h^{\text{AH}}(v) - U e^{\kappa v} + \frac{1}{\kappa} \dv{r_h^{\text{AH}}}{v},
	   	\end{equation}
	   where in the above equation we also take the black hole temperature to be almost constant over the timescale we are working in ($\sim$ order of the light crossing time). The surface gravity $\kappa$ is evaluated at the apparent horizon and $\dot{\kappa} \sim O(\varepsilon)$.\footnote{The cut-off surface could be very far away, and in general we do not expect the surface gravity corresponding to the location of island $\kappa_a$ to match with the surface gravity corresponding to point at the cut-off surface $\kappa_c$.}
	\end{enumerate}
	In this approximation we can linearize the blackening factor about its zero
	\begin{equation} \label{eq:blackening_linearized}
		f(r(U, v), v) \approx \kappa \left( r - r_h^{\text{AH}}(v) \right) = -\kappa U e^{\kappa v} + O(\varepsilon)
	\end{equation}
	So the generalized entropy in the Unruh state becomes
	\begin{equation} \label{eq:Sgen_U_approx}
		S_{\text{gen}}^{\text{U}}(U_a, v_a; U_c, v_c) \approx \frac{\lambda^2 a^2}{4G_N} + \frac{c}{6} \log \frac{\abs{v_a - v_c}}{\eta} + \frac{c}{6} \log\frac{U_a - U_c}{\eta} + \frac{c}{12} \kappa_a v_a + \frac{c}{6} \rho_c + \varepsilon \frac{c}{6} h(a),
	\end{equation}
	where we have used the approximation for $f$ in \eqref{eq:blackening_linearized} to simplify $\rho(U, v) \approx \frac{1}{2} \kappa v + \varepsilon h(r(U, v))$ and $a \equiv r(U_a, v_a)$. We extremize this generalized entropy and keep track of terms up to $O(1)$
	\begin{align}
		\pd_U{S_{\text{gen}}^{\text{U}}}(U, v) \Big|_{(U_a, v_a)} = 0 &\implies \frac{2\lambda^2 a}{\varepsilon} (-e^{\kappa_a v_a}) + \frac{1}{U_a - U_c} = 0 \label{eq:evap_U_extremization} \\
		\pd_v{S_{\text{gen}}^{\text{U}}(U, v)} \Big|_{(U_a, v_a)} = 0 &\implies \frac{4\lambda^2 a}{\varepsilon} \left( \dot{r}_h^{\text{AH}}(v_a) - \kappa_a U_a e^{\kappa_a v_a} \right) + \kappa_a + \frac{2}{v_a - v_c} = 0, \label{eq:evap_v_extremization}
	\end{align}
	where $\dot{r}_h^{\text{AH}}(v) = - \frac{\varepsilon}{16 \lambda^2 r_0^2}$ is just a constant at $O(\varepsilon)$. Simplifying these equations gives $U_a$ in terms of $U_c$ as
	\begin{equation} \label{eq:evap_island_loc_U}
		U_a = U_c \frac{4 \lambda^2 a\, \dot{r}_h^{\text{AH}}(v_a) (v_a - v_c) + \varepsilon \kappa_a (v_a - v_c) + 2\varepsilon}{4 \lambda^2 a\, \dot{r}_h^{\text{AH}}(v_a) (v_a - v_c) - \varepsilon \kappa_a (v_a - v_c) + 2\varepsilon},
	\end{equation}
	Clearly the fraction is $O(1)$ to leading order.

	Using \eqref{eq:radial_approx} and \eqref{eq:evap_v_extremization} we can find the location of the QES as
	\begin{equation} \label{eq:evap_QES}
		a = r_h(v_a) - \frac{\varepsilon}{8\lambda^2 r_0} \frac{4 + \kappa_0 (v_a - v_c)}{\kappa_0 (v_a - v_c)},
	\end{equation}
	where we have approximated $r_h^{\text{AH}}(v_a) \simeq r_0$ and $\kappa_a \simeq \kappa_0$ in the term already at $O(\varepsilon)$. Unlike the island for the eternal black hole, in this case the island is located at $O(cG_N)$ from the quantum-corrected event horizon and inside it for $\kappa_0 (v_a - v_c) < -4$. For $-4 < \kappa_0 (v_a - v_c) < 0$, the QES lies beyond the horizon.

	Plugging in the value of $U_a - U_c$ from \eqref{eq:evap_U_extremization}, the generalized entropy in the Unruh state is given by
	\begin{equation} \label{eq:Sgen_U_extremized}
		S_{\text{gen}}^U(v_a; v_c) \approx \frac{\lambda^2 a^2}{4G_N} - \frac{c}{6} \log \frac{a}{r_0} + \frac{c}{6} \log \frac{v_c - v_a}{\eta} - \frac{c}{12} \kappa_a v_a + \frac{c}{6} \rho_c + \frac{c}{6} \log \frac{\varepsilon}{2\lambda^2 r_0 \eta} + \varepsilon h(a).
	\end{equation}
	We note that initially the Bekenstein-Hawking term given by $\frac{\lambda^2 r_0^2}{4G_N}$ will be the dominant contribution to the entropy. Even though a closed form analytic expression for $v_a(v_c)$ is not available from the above extremization equations, to find the late time behaviour of the entropy we assume that $v_a - v_c = x$ where $x$ is some shift which does not depend on $v_c$ to the leading order in $O(\varepsilon)$. Then using \eqref{eq:evap_QES} in \eqref{eq:Sgen_U_extremized} we can write
	\begin{equation} \label{eq:evap_Sgen_approx_radial}
		\frac{\lambda^2 a^2}{4G_N} - \frac{c}{6} \log \frac{a}{r_0} \approx \frac{\lambda^2 r_0^2}{4G_N} - \frac{c \kappa_0 v_c}{24} + \ldots
	\end{equation}
	where ``$\ldots$'' represents terms either independent of $v_c$ or subleading $\sim O\left( \frac{\varepsilon}{\lambda^2 r_0^2} \right)$. Additionally from the term with the conformal factor evaluated at the point on the cut-off surface we find for late times \[ \rho_c \to \frac{1}{2} \kappa_0 v_c + \text{subleading terms}. \] Hence the late time behaviour of the radiation entropy in the Unruh state including the island contribution is
	\begin{equation} \label{eq:evap_Sgen_late_times}
		S_{\text{island}}^{\text{U}} \xrightarrow{\kappa_0 v_c \gg 1} \frac{\lambda^2 r_0^2}{4G_N} - \frac{c}{24} \kappa_0 v_c.
	\end{equation}
	Clearly, the island entropy is decreasing in the Unruh state, which is essentially different from the Hartle-Hawking case.
	
	\subsection{No-island case} \label{sec:evap_no_island}
	\begin{figure}[hptb]
		\centering
		\includegraphics{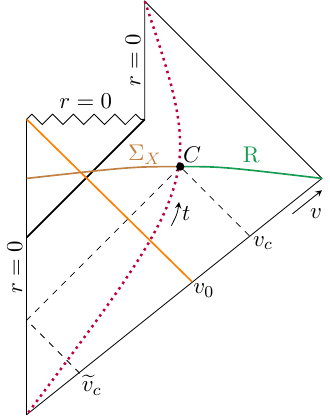}
		\caption{In the case of no-island, the entropy of the radiation region $S(\text{R})$ is same as the entropy of the complementary region $\Sigma_X$ and is encoded in the conformalon solution. The dashed line connecting $C$ to the $r = 0$ surface is the incoming ray $U = U_c$ traced back to the $r = 0$ surface.}
		\label{fig:evap_noisland}
	\end{figure}
	To find the no-island entropy, we first locate the image point of $\pd$R $= C$ with coordinates, say, $(\widetilde{U}_c, \widetilde{v}_c)$. This can be found using \eqref{eq:uin_U} by tracing the incoming null ray passing through $C$ all the way back to the boundary $r = 0$ to get
	\begin{equation} \label{eq:image_point}
		\begin{split}
			\widetilde{U}_c &= v_c - v_0 + 2r_0 \\
			\widetilde{v}_c &= U_c + v_0 - 2r_0.
		\end{split}
	\end{equation}
	The no-island entropy is then computed from \eqref{eq:no_island} as
	\begin{equation} \label{eq:rad_entropy_no_island_evap}
		S_{\text{no-island}}^{\text{U}} = \frac{c}{6} \left[ \rho(\widetilde{U}_c, \widetilde{v}_c) + \rho(U_c, v_v) + \log(\frac{v_c - v_0 + 2r_0 - U_c}{\eta})^2 \right] ,
	\end{equation}
	for late times the term $\rho(U_c, v_c) \to \frac{1}{2} \kappa_0 v_c$ and the other terms will be either $\log v_c$ or sub-leading in the $\varepsilon$ expansion. Thus
	\begin{equation} \label{eq:growing_Sgen_U}
		S_{\text{no-island}}^{\text{U}} \xrightarrow{\kappa_0 v_c \gg 1} \frac{c}{12} \kappa_0 v_c.
	\end{equation}
	In the Unruh state as well, not including the islands give an unbounded entropy growth for the radiation at late times.

	\subsection{Page time and Page curve} \label{sec:evap_page}
	\begin{figure}[hptb]
		\centering
		\includegraphics{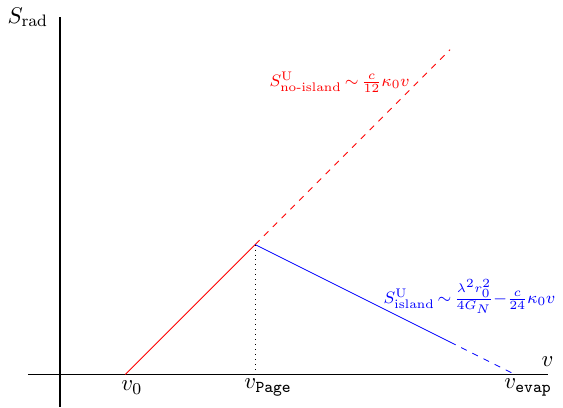}
		\caption{The Page curve for the radiation in the Unruh state within the linearized semi-classical approximation is indicated by the solid red and blue curves. Close to $v = v_{\text{evap}} \sim O(M_0^3)$ the semi-classical approximation can't be trusted (we denote this regime by the blue dashed line). The Page time is at $v_{\text{Page}} \sim \frac{G_N^2 M_0^3}{c \lambda^4}$ with $M_0 \gg 1$.}
		\label{fig:evap_Page}
	\end{figure}
	The fine-grained entropy of the radiation in the Unruh state is the minimum of the growing no-island entropy in \eqref{eq:rad_entropy_no_island_evap} or the falling island entropy \eqref{eq:evap_Sgen_late_times}. Initially, the no-island entropy is smaller but eventually at Page time the island entropy becomes smaller than the no-island case. The Page time can be computed from
	\begin{equation} \label{eq:Page_time_evap}
		\frac{\lambda^2 r_0^2}{4G_N} - \frac{c}{24r_0} v_{\text{Page}} \approx \frac{c}{12r_0} v_{\text{Page}} \implies v_{\text{Page}} \approx \frac{1024\pi^3 G_N^2 M_0^3}{c \lambda^4}.
	\end{equation}

	\section{Conclusion and outlook} \label{sec:discussion}
	In this paper we were able to show that the Wald-like prescription for generalized entropy in the island formula described in \cite{pedraza2021Wald, hirano2023WaldIsland} captures the corresponding behaviour of the quantum extremal surface and the Page curve in the DREH model, which approximates the s-wave sector of the 4D Einstein-Hilbert gravity. We dimensionally reduced the Einstein-Hilbert action to a 2D dilaton gravity model, and then studied the eternal black hole setup in the Hartle-Hawking state and the late time regime of the quasi-stationary evaporating black hole in the Unruh state. The quantum correction due to the matter ($N$ free fields) was incorporated via the Polyakov-Liouville term, and the combined gravity plus matter plus anomaly system was solved. The generalized entropy for the island in the backreacted geometry was obtained as the Noether-Wald charge of the total action: the dimensionally reduced Einstein-Hilbert term plus the 1-loop Polyakov-Liouville term. In particular, the conformalon solution in the appropriate vacuum accounts for the von Neumann entropy of the Hawking radiation and agrees with the Euclidean replica computation in quantum field theory (in curved spacetime) \cite{callan1994geoentropy, holzhey1994geoentropy, calabrese2004QFT, calabrese2009CFT}. 
	
	The results obtained here for the eternal black hole agree with the derivations based on island rule in \cite{djordjevic2022eternal} with the difference that we do not employ an explicit regularization scheme for the fine-grained entropy as done in \cite[eq.(75)]{djordjevic2022eternal}.\footnote{Our choices of normalization coefficients in the action and backreaction expansion parameter also differs from those in \cite{djordjevic2022eternal, djordevic2025collapse, djordevic2025evaporating}. We defined the 2D Newton's constant as $G_N = \frac{\lambda^2}{\pi} G_N^{(4)}$, as compared to $G_N = \lambda^2 G_N^{(4)}$ in \cite{djordjevic2022eternal}. The coefficient of the Polyakov-Liouville action in our case was taken to be $-\frac{c}{24\pi}$ whereas it is $-\frac{\hbar}{96\pi}$ in \cite{djordjevic2022eternal} -- while we took $N$ free massless scalar fields minimally coupled to the s-wave truncated gravity, in \cite{djordjevic2022eternal} the authors consider a single minimally coupled massless scalar as the matter sector. Finally, our expansion parameter is $\frac{2c G_N}{3}$ while in \cite{djordjevic2022eternal} it is $\frac{\hbar}{48\pi}$. We set $\hbar = 1$.} We also found that unlike in \cite{hashimoto2020Schwd} where the s-wave sector QES appears at order $(c G_N)^2$ away from the classical Schwarzschild radius, when backreaction is considered the QES is away from the classical Schwarzschild radius at the linear order in the backreaction parameter. We emphasize again that in this setup the Hawking radiation propagates towards the asymptotic infinity and we do not make any reference to a non-gravitating bath, unlike the AdS${}_2$ plus bath setup of \cite{penington2019EWR, penington2019replicaWH, almheiri2019BHIP, almheiri2019Page, almheiri2019replicaWH, almheiri2019island} where the bath collects the Hawking quanta. Moreover, unlike the AdS${}_2$ plus bath setup where the physical boundary time gave the natural time parameter with respect to which the no-island entropy grew linearly at late times, the parameter along the timelike cut-off surfaces shown in Figures \ref{fig:eternal_island} and \ref{fig:eternal_noisland} which separate the spacetime into the black hole region and radiation region give the time-dependence for the no-island entropy in our setup.
	
	For the evaporating black hole we found that the QES is located at $O(cG_N)$ away from the backreacted event horizon, however unlike the eternal case it can be within the event horizon or outside the horizon. This behaviour is different from what was reported for the Page curve for evaporating black hole in \cite{djordevic2025evaporating}. Another interesting feature of the evaporating case, which is crucially different from the eternal case is that the radial location of QES is manifestly time dependent. From \eqref{eq:eternal_QES} we see that given the cut-off surface $r = r_c \iff r_*(r_c) = c_*$, the location of the QES is uniquely fixed from the extremization conditions for the eternal black hole. However, from \eqref{eq:evap_U_extremization}-\eqref{eq:evap_QES} we notice that given $(U_c, v_c)$ we have a family of QES that remains parametrically close to the event horizon, unlike a single QES following a causal trajectory as in the eternal black hole.\footnote{A similar behaviour was noted for the QES in evaporating black holes for the non-minimally coupled DREH model in \cite{wu2023islands}. We thank Chih-Hung Wu for discussions and clarification on this point.}
	
	In this work we first reduced to the s-wave sector and then added the quantum correction in the 2D theory, with the matter sector taken to be minimally coupled to gravity. While this provides a good approximation to s-wave sector of the full 4D theory and in particular, to study the Hawking radiation which is dominated by the lowest angular momentum modes \cite{penington2019EWR}, it leaves out some important aspects of the 4D physics such as the appearance of greybody factors and vacuum polarization effects, the former being absent in the two-dimensional theory. Moreover, it is well known that the quantization procedure does not commute with the process of dimensional reduction \cite{frolov1999dimredanomaly}. These issues highlight several aspects along which the current work could be extended:

	\subsection*{Dimensional reduction of four-dimensional matter}
	While the gravitational part $I_{\text{DG}}$, of the action in \eqref{eq:total_action} had a proper origin in the 4D theory, namely the Einstein-Hilbert action in \eqref{eq:EH_action}, the matter action in \eqref{eq:classical_action} was somewhat ad hoc. We did not apply any dimensional reduction to a matter theory in 4D, and the matter sector was taken to be minimally-coupled which greatly simplified the equations of motion. In particular, the dilaton equation of motion was unchanged when the 1-loop effects were included. However, it has been pointed out in earlier works \cite{mukhanov1994HawkRad, balbinot2000vacpolarization, fabbri2005evaporation, wu2023islands} that if we start with a 4D matter action and dimensionally reduce it to two dimensions, we will have non-trivial matter-dilaton couplings of the form
	\begin{equation} \label{eq:nonmin_dilaton}
		I_{\text{m}} \sim - \int \dd^2 x\, \sqrt{-\det g}\, g^{ab}\, e^{-2\phi}\, \nabla_a \vec{\chi}\, \cdot \nabla_b \vec{\chi},
	\end{equation}
	varying which will give non-trivial dilaton-anomaly mixing terms in the equations of motion as well. So the dilaton equation of motion will not remain unaffected when the 1-loop effects are taken into account. This model has been well studied from different approaches, albeit with some incompatibilities. It is perhaps a better approximation to the quantum aspects of black holes in the s-wave sector compared to the Polyakov-Liouville theory describing the minimally-coupled matter sector, modelled as a CFT (in this work as $N$ massless scalars). The QES in different vacua have been studied for this non-minimal dilaton gravity model via the island formula in \cite{wu2023islands}. It would be interesting to apply the Wald-like entropy prescription to see if the conformalon solution still captures the von Neumann entropy of the matter sector modelling the Hawking radiation. We hope to report on this in an upcoming work.

	\subsection*{Reduction or Quantization: which first?}
	Since the reduction and the quantization procedures do not commute in general, it would be interesting to study the semi-classical Page curve for the quantum-corrected eternal and evaporating black hole in the 4D setting first and then verify if the s-wave sector matches with the results from dimensional reduction. Solving the full backreaction problem in 4D still remains an open problem, although in \cite{shafiee2022scSchwd, shafiee2026HawkRad} the quantum corrections to the Schwarzschild black hole solution was obtained incorporating both the vacuum polarization and conformal anomaly effects. The resulting black hole has two horizons, and the end point of the evaporation is an extremal remnant with zero temperature. This suggests a much richer black hole thermodynamics for the 4D setup, resulting in a more non-trivial Page curve, which was not studied in the works of Shafiee et. al. It would be interesting to study the Page curve for this model and compare the s-wave sector to our results.

	An important aspect of the 4D quantization \cite{birrell1984, buchbinder1992effaction} is that on top of the conformal anomaly contribution \cite{christensen1977anomaly}, there is also the vacuum polarization contribution \cite{candelas1980vacpolarization, howard1984quantumSET, balbinot1999vacpolarization, balbinot2000vacpolarization, bardeen2017scSET, bardeen2018scSET}. Assuming that the Hawking radiation is dominated by the massless sector, particularly the conformally coupled massless sector \cite{hirano2023WaldIsland}, the backreaction can be incorporated as the conformally-coupled matter plus the conformal anomaly action. The latter is the Riegert action \cite{riegert1984anomaly} which can be cast into a local form similar to the Polyakov-Liouville term in 2D, using two auxiliary fields \cite{mottola2006anomaly, mottola2016anomaly}. Then a similar analysis done in the Wald-like prescription should be applicable to the 4D theory as well \cite{hirano2023WaldIsland}. However, the full 1-loop effective action also has contributions from the vacuum polarization effects which can be organized into an expansion in generalised curvature invariants, where at each order of the curvature expansion we need to introduce an auxiliary field to render the action local \cite{barvinsky1985SchwingerDeWitt, barvinsky1987covariant1, barvinsky1990covarinat2}. The Wald-like prescription could, in principle, be used to incorporate the vacuum polarization effects if we choose to work up to a particular order in the curvature expansion in the effective action. This should be contrasted with the two-dimensional case where the conformal anomaly completely captures the 1-loop correction of the matter via the Polyakov-Liouville term \cite[Footnote 13]{hirano2023WaldIsland}.

	\subsection*{Euclidean replica wormholes and Lorentzian Wald-like entropy}
	Earlier works \cite{pedraza2021Wald, pedraza2021microcanonical, hirano2023WaldIsland} suggested that island computations, based on the Euclidean replica wormhole, in JT and RST gravity models agree with the extremization of the generalized entropy obtained as a Lorentzian Noether-Wald charge. In this work, we extended the Lorentzian analysis to the black hole solutions in the DREH model and showed that the results are indeed in agreement with the island formula computations of \cite{djordjevic2022eternal}. We suspect these agreements are not accidental, but instances of a more general statement: for any diffeomorphism-invariant theory, the generalized entropy extremized in the island formula is itself the Wald–Noether charge of the combined gravity plus matter theory, evaluated at the location of cosmic brane on the (orbifolded) $n$-fold replica manifold, in the $n \to 1$ limit. This has been hinted at in earlier works \cite{dong2013HEE, camps2013Sgen, dong2017entropy} and it would be worth making this identification precise.
	
	\acknowledgments
	We would like to thank Anupam A.H, Shivan K. Sharma, and Ronak Soni for many useful discussions. We also thank Moslem Shafiee, and Chih-Hung Wu for helpful correspondence and clarifications.
	
	\appendix
	\addtocontents{toc}{\protect\setcounter{tocdepth}{1}}
	\section{Black hole solutions in the DREH model} \label{sec:bh_soln}
	In this appendix we discuss the to the quantum-corrected black hole solutions in the DREH model. We will only consider the vacuum solution to the quantum-corrected equations of motion in \eqref{eq:metric_eom_quantum}. 

	\subsection{Eternal black holes} \label{sec:eternal_soln}	
	We start with the metric ansatz in the Schwarzschild chart\footnote{We could start with a more generic metric of the form $-f_0(r) \dd t^2 + g_0(r) \dd r^2$, but the $rr$ and $tt$ equations of motion will imply that $f_0 g_0 = \text{constant}$. By a suitable rescaling of time coordinate we can set this constant to 1.}
	\begin{equation} \label{eq:eternal_BH_Schwd}
		\dd s^2 = -f_0(r) \dd t^2 +  \frac{\dd r^2}{f_0(r)},
	\end{equation}
	with $r$ being the radial coordinate given in \eqref{eq:radial_dilaton}. The Ricci scalar is given by $R = -f''_0(r)$, where $\ '\ \equiv \dv{r}$. The dilaton equation \eqref{eq:dilaton_eom_cl} then becomes
	\begin{equation} \label{eq:eternal_dilaton_cl_simple}
		f''_0(r) - \frac{2}{r} f'_0(r) = 0.
	\end{equation}
	Also, the $g_{rr}$ equation of motion is
	\begin{equation} \label{eq:grr_eom_cl}
		f_0(r) = 1 - r f'_0(r).
	\end{equation}
	Together \eqref{eq:eternal_dilaton_cl_simple} and \eqref{eq:grr_eom_cl} implies
	\begin{equation} \label{eq:blackening_factor_cl}
		f_0(r) = 1 - \frac{A}{r},
	\end{equation}
	where the constant $A$ can be fixed by relating it to the black hole mass $M$ in 4D
	 \begin{equation} \label{eq:fixing_mass_cl}
		A = 2M G_N^{(4)} = \frac{8\pi G_N M}{\lambda^2} \equiv r_0,
	\end{equation}
	where $r = r_0$ gives the location of the horizon of the classical black hole whose metric is
	\begin{equation} \label{eq:Schwd_cl}
		\dd s^2 = -\left( 1 - \frac{r_0}{r} \right) \dd t^2 + \left( 1 - \frac{r_0}{r} \right)^{-1} \dd r^2 .
	\end{equation}
	
	To the first order in $\varepsilon$ defined in \eqref{eq:expansion_parameter},  we will find an eternal black hole solution incorporating the 1-loop correction. The backreaction generally makes the semi-classical metric depart from the classical ``Schwarzschild form'', and we need two functions to incorporate the correction to the blackening factor $(m(r))$ and to the redshift/lapse factor $(h(r))$. We will work with the following metric ansatz
	\begin{equation} \label{eq:metric_ansatz_sc}
		\dd s^2 = -f(r) e^{2\varepsilon h(r)} \dd t^2 + \frac{\dd r^2}{f(r)},
	\end{equation}
	with the blackening factor
	\begin{equation} \label{eq:blackening_factor_sc}
		f(r) = f_0(r) + \varepsilon \frac{m(r)}{r}.
	\end{equation}
	The location of the horizon, $r_h = r_0 + \varepsilon r_1$, to the linear order is given by
	\begin{equation} \label{eq:horizon_sc}
		f(r_h) = 1 - \frac{r_0}{r_h} + \varepsilon \frac{m(r_h)}{r_h} = 0 \implies r_1 = -m(r_0).
	\end{equation}

	The semi-classical $tt$ and $rr$ equations of motion at the linearized order are obtained from \eqref{eq:metric_eom_quantum} 
	\begin{align}
		\varepsilon f_0(r) m'(r) &= -\frac{8\pi G_N}{\lambda^2} \ev{\T_{tt}} \label{eq:tt_eom_sc}, \\
		2\varepsilon r f_0(r) h'(r) &= \frac{8\pi G_N}{\lambda^2} \left( f_0(r) \ev{\T_{rr}} + \frac{\ev{\T_{tt}}}{f_0(r)} \label{eq:rr_eom_sc} \right)\, ,
	\end{align}
	where the stress tensor expectation values are evaluated in the Hartle-Hawking state. At this order, the dilaton equation is trivially satisfied and will not play a role in further analysis. Moreover, we can use the classical metric at this order to solve for the conformalon in \eqref{eq:conformalon_eqn}, which enters the stress tensor expectation values above. We will solve for the conformalon in the Hartle-Hawking state, which then gives the generalized entropy in \eqref{eq:Sgen_Wald} (cf. Section \ref{sec:eternal}). This is conveniently done in the Kruskal-Szekeres coordinates which we denote as $(x^+, x^-)$. In these coordinates, the Hartle-Hawking state is the vacuum state defined with respect to the positive-frequency modes in $x^\pm$ coordinates.

	We introduce the tortoise coordinate as
	\begin{equation} \label{eq:tortoise_coord}
		\dd r_* = \frac{\dd r}{f(r) e^{\varepsilon h(r)}} \implies r_* = \int \dd r\, \frac{e^{-\varepsilon h(r)}}{f(r)}.
	\end{equation}
	Then Kruskal-Szekeres coordinates are given by
	\begin{equation} \label{eq:KS_coord}
		x^\pm = \pm \frac{1}{\kappa} e^{\pm \kappa (t \pm r_*)},
	\end{equation}
	where $\kappa$ is the surface gravity
	\begin{equation} \label{eq:surface_grav}
		\kappa = \frac{1}{2} \pd_r(f(r) e^{\varepsilon h(r)}) \bigg|_{r = r_h} = \frac{1}{2r_0} \left[ 1 + \frac{\varepsilon}{\lambda^2 r_0^2} \left( h(r_0) + m'(r_0) + \frac{m(r_0)}{r_0} \right) + O(\varepsilon^2) \right].
	\end{equation}
	In these coordinates, the metric takes a conformal form
	\begin{equation} \label{eq:metric_sc_KS}
		\dd s^2 = -e^{2\rho(x)} \dd x^+ \dd x^-,
	\end{equation}
	with the conformal factor given by
	\begin{equation} \label{eq:conformal_factor_eternal}
		\rho(x) = \frac{1}{2} \log f + \varepsilon h - \kappa r_*.
	\end{equation}
	
	The conformalon equation particularly simplifies in these coordinates to
	\begin{equation} \label{eq:conformalon_eq_KS}
		\pd_+ \pd_- (\psi + \rho) = 0.
	\end{equation}
	If we denote the solution for the homogeneous part as $\psi_\pm(x^\pm)$, then the conformalon solution is
	\begin{equation} \label{eq:conformal_soln_KS}
		\psi(x) = -\rho(x) + \psi_+(x^+) + \psi_-(x^-).
	\end{equation}
	The functions $\psi_\pm(x^\pm)$ characterize the state we choose to work with. In particular, the combination
	\begin{equation} \label{eq:t_pm_defn}
		t_\pm(x^\pm) = \pd_\pm^2 \psi_\pm(x^\pm) - (\pd_\pm \psi_\pm(x^\pm))^2,
	\end{equation}
	will be proportional to the expectation value of the normal ordered part of the stress tensor operator \cite{fabbri2005evaporation}. The quantum correction to the stress tensor in \eqref{eq:1loop_EM} takes a simple form in the conformal gauge
	\begin{equation} \label{eq:1loop_EM_null}
		\begin{split}
	   	\ev{\T_{\pm \pm}} &= \frac{\varepsilon}{8\pi G_N} \left[ \pd_\pm^2 \rho - (\pd_\pm \rho)^2 - t_\pm \right], \\
	   	\ev{\T_{+-}} &= - \frac{\varepsilon}{8\pi G_N} \pd_+ \pd_- \rho.
		\end{split}
	\end{equation}

	For a given choice of vacuum $\ket{0_y}$ with respect to the coordinate frame $y$, the stress tensor operator can be split as \cite{deWitt1975QFTCS, davies1976EMTdecompose}
	\begin{equation} \label{eq:EM_op_split}
	    T_{ab}(y) =\ \normT{y}  + \bra{0_y} T_{ab}(y) \ket{0_y}.
	\end{equation}
	Then by definition \[ \ev{\normT{y}}{0_y} = 0. \] In particular, for the Hartle-Hawking state we have
	\begin{equation} \label{eq:HH_vac}
	    \ev{\normT{x}}{HH} = 0 \implies t_\pm(x^\pm) = 0.
	\end{equation}
	Since RHS in \eqref{eq:1loop_EM_null} is already at $O(\varepsilon)$ we can expand $\rho$ to the classical order $O(\varepsilon^0)$ i.e.,
	\begin{equation} \label{eq:rho_derivative_eternal}
		\pd_\pm \rho = \frac{1}{2f_0} \pd_\pm f_0 - \kappa \pd_\pm r_* = \frac{1}{4 \kappa x^\pm} \left( \frac{r_0}{r^2} - \frac{1}{r_0} \right),
	\end{equation}
	and simplify for the stress tensor components as
	\begin{equation} \label{eq:1loop_EM_null_simplified}
		\begin{split}
			\ev{\T_{\pm \pm}}{HH} &= \frac{\varepsilon}{32\pi G_N} \frac{(r - r_0)^2}{(x^\pm)^2 r^4} \left( r^2 + 2r_0 r + 3 r_0^2 \right), \\
			\ev{\T_{+-}}{HH} &= \frac{\varepsilon}{8\pi G_N} \frac{r_0^3 (r - r_0)}{x^+ x^- r^4}.
		\end{split}
	\end{equation}
	Transforming back to the Schwarzschild chart we have
	\begin{equation} \label{eq:1lopp_EM_simplified}
		\begin{split}
			\ev{\T_{rr}}{HH} &= \pdv{x^a}{r} \pdv{x^b}{r} \ev{\T_{ab}}{HH} = \frac{\varepsilon}{64\pi G_N} \frac{1}{f_0^2} \left( \frac{1}{r_0^2} - \frac{r_0^2}{r^4} \right), \\
			\ev{\T_{tt}}{HH} &= \pdv{x^a}{t} \pdv{x^b}{t} \ev{\T_{ab}}{HH} = \frac{\varepsilon}{64\pi G_N} \left( \frac{1}{r_0^2} - \frac{8r_0}{r^3} + \frac{7r_0^2}{r^4} \right) .
		\end{split}
	\end{equation}
	The quantum correction to the metric is then easily solved for using \eqref{eq:tt_eom_sc} and \eqref{eq:rr_eom_sc} as
	\begin{equation} \label{eq:m_h_soln_eternal}
		\begin{split}
			m(r) &= \frac{1}{8\lambda^2 r_0} \left( -\frac{7}{2} \frac{r_0^2}{r^2} + \frac{r_0}{r} - \log r - \frac{r}{r_0} \right) + D_1, \\
			h(r) &= \frac{1}{8\lambda^2 r_0^2} \left( -\frac{3}{2} \frac{r_0^2}{r^2} - 2 \frac{r_0}{r} + \log r \right) + D_2.
		\end{split}
	\end{equation}
	The constant $D_1, D_2$ can be fixed by introducing a large-distance length scale $L$, such that $m(r), h(r) \to 0$ as $r \to L$ and the length scale is taken to $\infty$.\footnote{Since in the metric only $\frac{m(r)}{r}$ appears, as $r \to \infty$, $f(r)$ is well behaved and tends to 1 even without explicitly introducing the large distance cut-off.} Then 
	\begin{equation} \label{eq:constants}
		\begin{split}
			D_1 &= \frac{1}{8\lambda^2 r_0} \left( \frac{L}{r_0} + \log L \right) \\
			D_2 &= -\frac{1}{8\lambda^2 r_0^2} \log L
		\end{split}	
	\end{equation}
	The horizon radius from \eqref{eq:horizon_sc} and the surface gravity \eqref{eq:surface_grav} are then corrected as
	\begin{align}
		r_h &= r_0 + \frac{\varepsilon}{8\lambda^2 r_0} \left( \frac{7}{2} - \frac{L}{r_0} + \log \frac{r_0}{L} \right), \label{eq:corrected_horizon_eternal} \\
		\kappa &= \frac{1}{2r_0} \left[ 1 + \frac{\varepsilon}{8\lambda^2 r_0^3} (L - 3r_0 ) \right]. \label{eq:surface_grav_sc_eternal}
	\end{align}
	Thus the black hole temperature, which is proportional to the surface gravity, also receives quantum corrections.

	We conclude this section with one final result we will use about the tortoise coordinate close to the backreacted horizon. Expanding the integrand in \eqref{eq:tortoise_coord} close to $r_h$ and using the definition of surface gravity, we have
	\begin{equation} \label{eq:tortoise_approx}
		\begin{split}
			r_* &\approx \frac{1}{2\kappa} \int \dd r\, \left[ \frac{1}{r - r_h} - \left( \frac{1}{2} \frac{f''(r_h)}{f'(r_h)} + \varepsilon\, h'(r_h) \right) + O(r - r_h) \right] \\
			&= \frac{1}{2\kappa} \left[ \frac{r}{r_h} + \log(\frac{r}{r_h} - 1) + \varepsilon \underbrace{\frac{3r}{8\lambda^2 r_h^3}}_{g(r)} \right].
		\end{split}
	\end{equation}

	\subsection{Evaporating black hole} \label{sec:evaporating_soln}
	We now describe the quantum-corrected evaporating black hole solution in the DREH model. The classical solution in this case is given by the Vaidya metric in the ingoing Eddington-Finkelstein coordinate $v$ as
	\begin{equation} \label{eq:classical_evaporating}
		\dd s^2 = -f_0(r, v) \dd v^2 + 2 \dd v \dd r, \qq{with} f_0(r, v) = 1 - \frac{r_0(v)}{r} = 1 - \frac{8\pi G_N M(v)}{\lambda^2 r},
	\end{equation}
	and the mass function is given by
	\begin{equation} \label{eq:mass_function}
		M(v) = M_0 \Theta(v - v_0).
	\end{equation}
	where $\Theta$ is the Heaviside step function. The Ricci scalar for this metric is
	\begin{equation} \label{eq:evap_Ricci_cl}
		R = f''_0(r, v) = \frac{2 r_0}{r^3}.
	\end{equation}
	where $r_0 = \frac{8\pi G_N M_0}{\lambda^2}$
	
	The evaporating black hole formed from collapse corresponds to the $\ket{in}$ state of the quantum fields \cite{davies1976evaporating, hiscock1980evaporating, hiscock1981evaporating}, in which case the expectation values of the stress tensor components are crucially time-dependent. We will not be interested in modelling the specific details of the collapse and the initial transients after the black hole formation, and instead only consider the late time regime in which the $\ket{in}$ state coincides with the Unruh state $\ket{U}$. In the latter case, the explicit time-dependence of the stress tensor expectation values vanishes \cite{fabbri2005evaporation}. Before proceeding with the details of the backreaction we summarize the similarities and differences between these two states (in Table \ref{tab:in_Unruh}) to justify why in the quasi-stationary approximation both states give the same backreaction physics.

	\begin{table}[htbp]
		\centering
		\caption{Comparison of the $in$-state and the Unruh state}
		\label{tab:in_Unruh}
		\begin{tabular}{|>{\centering\arraybackslash}m{0.2\textwidth}|>{\centering\arraybackslash}m{0.35\textwidth}|>{\centering\arraybackslash}m{0.35\textwidth}|}
			\hline
			Property & $in$ state, $\ket{in}$ & Unruh state, $\ket{U}$ \\
			\hline
			$\ev{T_{vv}}$ at $\scri^-$ & 0 & 0 \\
			\hline
			$\ev{T_{uu}}$ at $\scri^+$ & 0 $\to$ transient phase $\to$ Hawking flux & Constant Hawking flux \\
			\hline
			Regularity at future horizon & Yes & Yes \\
			\hline
			Has a formation phase & Yes & No, knows nothing about any formation event in the past \\
			\hline
			Thunderpop & Yes & No \\
			\hline
			Applicability & Black hole formation from collapsing matter and eventual evaporation & Model steady-state Hawking flux and evaporation beyond the collapse phase \\
			\hline
		\end{tabular}
	\end{table}

	We will incorporate the 1-loop effects and study the linearized backreaction to the metric in \eqref{eq:classical_evaporating} within the quasi-stationary approximation. We start with the following ansatz for the backreacted metric
	\begin{equation} \label{eq:evap_metric_cl}
		\dd s^2 = -f(r, v) e^{2\varepsilon h(r, v)} \dd v^2 + 2e^{\varepsilon h(r, v)} \dd v \dd r,
	\end{equation}
	where
	\begin{equation} \label{eq:blackening_factor_evap}
		f(r, v) = 1 - \Theta(v - v_0) \left( \frac{8\pi G_N M_0}{\lambda^2 r} - \varepsilon \frac{m(r, v)}{r} \right).
	\end{equation}
	
	The semi-classical metric equations of motion at the linearized order are obtained from \eqref{eq:metric_eom_quantum} as 
	\begin{align}
		2\varepsilon r h'(r, v) &= \frac{8\pi G_N}{\lambda^2} \ev{\T_{rr}}, \label{eq:evap_rr_eom_sc} \\
		\varepsilon m'(r, v) &= \frac{8\pi G_N}{\lambda^2} \ev{\T_{vr}}, \label{eq:evap_vr_eom_sc} \\
		\varepsilon \dot{m}(r, v) &= - \frac{8\pi G_N}{\lambda^2} \left[ f_0(r, v) \ev{\T_{vr}} + \ev{\T_{vv}} \right], \label{eq:evap_vv_eom_sc}
	\end{align}
	where $\ \dot{}\ \equiv \pd_v$ and $\ '\ \equiv \pd_r$. The stress tensor expectation values on the RHS are evaluated in the Unruh vacuum. To compute the RHS above we need the expression for the conformalon in the Unruh vacuum. Since we are already working at $O(\varepsilon)$, we can use the classical metric at this order to solve for the conformalon in \eqref{eq:conformalon_eqn}. Moreover, it will be convenient to work with both the Eddington-Finkelstein coordinates which are given by
	\begin{equation} \label{eq:EF_coords}
		v,\ u = v - 2r_*,\ \text{with}\ \dv{r_*}{r} = \frac{1}{f(r, v)} e^{-\varepsilon h(r,v)},\ \text{for}\ v > v_0, 
	\end{equation}
	where $v = v_0$ is the surface of formation, and the metric takes the following form beyond it, (see Figure \ref{fig:evaporating_bh})
	\begin{equation} \label{eq:evap_metric_EF}
		\dd s_{\text{out}}^2 = -f(r(u_{\text{out}}, v), v) e^{2\varepsilon h(r(u_{\text{out}}, v), v)} \dd u_{\text{out}} \dd v \equiv -e^{2\rho(r(u_{\text{out}}, v), v)} \dd u_{\text{out}} \dd v.
	\end{equation}
	This is valid only in the patch $v_{\text{evap}} > v > v_0$, which is the backreacted black hole geometry.\footnote{Here we take the black hole to completely evaporate. Although we do not expect that the linear order analysis that we do in this work is sufficient to capture the complete evaporation of the black hole. In fact, the semi-classical treatment breaks down close to the endpoint of the evaporation.} We will denote the location of the event horizon which will now be dependent on $v$ as $r_h(v)$ and the location of the apparent horizon as $r_h^{\text{AH}}(v)$, such that $f(r_h^{\text{AH}}, v) = 0$, pointwise in $v$. This implies
	\begin{equation} \label{eq:AH}
		r_h^{\text{AH}}(v) = r_0 - \varepsilon m(r_0, v),
	\end{equation}
	The event horizon is defined teleologically, as the boundary of the causal past of future null infinity, in particular it will be the particular outgoing null geodesic that neither escapes to $\scri^+$ nor falls into the singularity. The outgoing null rays satisfy
	 \begin{equation} \label{eq:null_rays_out}
		 \dv{r}{v} = \frac{1}{2} \left[ f_0(r, v) + \varepsilon \frac{m(r, v)}{r} \right] e^{\varepsilon h(r, v)}.
	\end{equation}
	Suppose the candidate event horizon lies at $r_h(v) = r_0 + \varepsilon r_1(v)$, for $v > v_0$, . Then the quantity in the square brackets will be at $O(\varepsilon)$ and we can ignore the exponential factor containing $h(r, v)$ at the linear order. This give
	\begin{equation} \label{eq:EH_perturb_ODE}
		\dot{r}_1(v) - \kappa_0  r_1(v) = \kappa_0 m(r_0, v),
	\end{equation}
	where $\kappa_0 = \frac{1}{2r_0}$ is the classical surface gravity. The homogeneous solution for this equation is given by $e^{\kappa_0 v}$, thus
	\begin{equation} \label{eq:EH_perturb}
		r_1(v) = -\kappa_0 \int_v^\infty \dd v'\, m(r_0, v') e^{-\kappa_0(v' - v)}
	\end{equation}

	For $v < v_0$, the metric takes the Minkowski form given by
	\begin{equation} \label{eq:Mink_region_metric}
		\dd s_{\text{in}}^2 = - \dd u_{\text{in}} \dd v,
	\end{equation}
	with $u_{\text{in}}, v$ being the null Minkowski coordinates. We have the following relation for the coordinate $u$
	\begin{equation} \label{eq:outgoing_EF}
		u = \begin{cases} u_{\text{in}} = v - 2r, & v < v_0 \\ u_{\text{out}} = v - 2r_*, & v > v_0 \end{cases}
	\end{equation}
	and from the matching condition at $v = v_0$ we require continuity of the dilaton on both sides of $v = v_0$. This implies
	 \begin{equation} \label{eq:junction_condition}
		 r(u_{\text{in}}, v_0) = r(u_{\text{out}}, v_0).
	 \end{equation}
	 To solve this relation for $u_{\text{in}}$ in terms of $u_{\text{out}}$, we first need to invert $r_*(r)$ in terms of $r$ resulting in
	 \begin{equation} \label{eq:r_in_tortoise}
	 	r = r_0 + W(e^{\frac{r_*}{r_0} - 1})
	 \end{equation}
	 where $W(\cdot)$ is the Lambert-W function. Then from \eqref{eq:outgoing_EF} we get 
	\begin{equation} \label{eq:uin_uout}
		u_{\text{in}} = v_0 - 2r_0 - 2r_0 W\left( e^{-1 + \kappa_0 v_0 - \kappa_0 u_{\text{out}}} \right),
	\end{equation}
	where $\kappa_0 = \frac{1}{2r_0}$ is the classical surface gravity. For large $u_{\text{out}}$ we have
	\begin{equation} \label{eq:uin_U}
		u_{\text{in}} \approx v_0 - 2r_0 - \frac{1}{\kappa_0} e^{-\kappa_0 u_{\text{out}}} = v_0 - 2r_0 + U,
	\end{equation}
	where $U$ is the Kruskal coordinate with
	\begin{equation} \label{eq:Kruskal-Szekeres}
		U = -\frac{1}{\kappa_0} e^{-\kappa_0 u_{\text{out}}},
	\end{equation}
	Note that the coordinates $(u_{\text{in}}, v)$ are defined everywhere and the $\ket{in}$ state is precisely the vacuum state with respect to the positive frequency modes in these coordinates. Moreover, from the above relation, we see that the $\ket{in}$ state actually matches with the Kruskal vacuum $\ket{U}$ defined with respect to the positive frequency modes in $(U, v)$. This relation further supports our earlier claim that in the quasi-stationary approximation both states give the same physics in the late time regime. Below we will just use $u$ to denote the outgoing Eddington-Finkelstein coordinate.
	
	In the $(u, v)$ coordinates, for $v > v_0$, the conformalon equation simplifies to
	\begin{equation} \label{eq:evap_conformalon_eqn}
		\pd_u \pd_v (\psi(u, v) + \rho(u, v)) = 0.
	\end{equation}
	The solution for $\psi$ is given by
	\begin{equation} \label{eq:psi_particular}
		\psi(u, v) = -\rho(u, v) + \psi_1(v) + \psi_2(u).
	\end{equation}
	The complete solution for the conformalon has two homogeneous parts $\psi_1(v)$ and $\psi_2(u(r, v))$ which are fixed by the choice of the quantum state. We need to specify additional information such as $\ev{T_{vv}} = 0$ at $\scri^-$ for the Unruh/in state to fix $\psi_1(v)$, similar to what we had for the Hartle-Hawking state in \eqref{eq:HH_vac}. To fix $\psi_2(u)$ we will use the regularity at the (future) horizon with respect to the affine coordinate $U$. Similar to \eqref{eq:1loop_EM_null} the quantum correction to the stress tensor in \eqref{eq:1loop_EM} takes the form
	\begin{equation} \label{eq:1loop_EM_null_evap}
		\begin{split}
	   	\ev{\T_{vv}}_{(u,v)} &= \frac{\varepsilon}{8\pi G_N} \left[ \pd_v^2 \rho - (\pd_v \rho)^2 - t_v(v) \right], \\
	   	\ev{\T_{uu}} &= \frac{\varepsilon}{8\pi G_N} \left[ \pd_u^2 \rho - (\pd_u \rho)^2 - t_u(u) \right], \\
		\ev{\T_{uv}} &= - \frac{\varepsilon}{8\pi G_N} \pd_u \pd_v \rho,
		\end{split}
	\end{equation}
	where $\ev{\T_{vv}}_{(u,v)}$ indicates the expectation value in the $u, v$ chart, which is different from $\ev{\T_{vv}}$ in \eqref{eq:evap_vv_eom_sc}. The normal ordered part $t_v(v)$ is given as
	\begin{equation} \label{eq:t_v}
			t_{v} = \pd_v^2 \psi_2(v) - (\pd_v \psi_2(v))^2 = \ev{\normTchi[vv]{v}}{in} = \ev{\normTchi[vv]{v}}{U} = 0,
	\end{equation}
	and the last equality holds everywhere in $v$ because $v$ is an affine, horizon-regular coordinate along its entire ingoing null congruence. Similarly for $t_u(u)$ we have
	\begin{equation} \label{eq:t_u}
		\begin{split}
			t_{u} &= \pd_u^2 \psi_1(u) - (\pd_u \psi_1(u))^2 = \ev{\normTchi[uu]{u}}{in} = \ev{\normTchi[uu]{u}}{U} \\ &= \left( \pdv{U}{u} \right)^2 \ev{\normTchi[UU]{U}}{U} + \frac{1}{2} \{U, u\} = -\frac{1}{4} \kappa_0^2 = -\frac{1}{16 r_0^2},
		\end{split}
	\end{equation}
	where the first term vanishes due to the fact that $t_U(U) = 0$ in the Unruh state, and $\{U, u\}$ denotes the Schwarzian derivative of $U$ with respect to $u$.
	
	The stress tensor expectation values are then given by
	\begin{equation} \label{eq:1loop_EM_null_evap_simplified}
		\begin{split}
			\ev{\T_{uu}}{U} &= \frac{\varepsilon}{32\pi G_N} \left[ -\frac{r_0}{r^3} + \frac{3r_0^2}{4r^4} + \kappa_0^2 \right], \\
			\ev{\T_{uv}}{U} &= -\frac{\varepsilon}{32\pi G_N} \left[ \frac{r_0}{r^3} - \frac{r_0^2}{r^4} \right], \\
			\ev{\T_{vv}}{U}_{(u,v)} &= \frac{\varepsilon}{32\pi G_N} \left[ -\frac{r_0}{r^3} + \frac{3r_0^2}{4r^4} \right],
		\end{split}
	\end{equation}
	In obtaining the above expression we used the following relations 
	\begin{equation} \label{eq:rho_derivative_evap}
		\begin{split}
			\pd_u \rho(u, v) \big|_v &= -\frac{r_0}{4r^2}, \\
			\pd_v \rho(u, v) \big|_u &= \frac{r_0}{4r^2}.
		\end{split}
	\end{equation}

	Transforming back to the ingoing Eddington-Finkelstein chart we have
	\begin{equation} \label{eq:1loop_EM_null_evap_EF_simplified_rr}
		\begin{split}
			\ev{\T_{rr}}{U} &= \left( \pdv{u}{r} \right)^2 \ev{\T_{uu}}{U} + \pdv{u}{r} \pdv{v}{r} \ev{\T_{uv}}{U} + \left( \pdv{v}{r} \right)^2 \ev{\T_{vv}}{U}_{(u,v)} \\
			&= \frac{4}{f_0^2} \ev{\T_{uu}}{U} = \frac{\varepsilon}{32\pi G_N r_0^2} \left[ 1 + \frac{2r_0}{r} + \frac{3r_0^2}{r^2} \right].
		\end{split}
	\end{equation}
	Similarly for the other two components we get
	\begin{equation} \label{eq:1loop_EM_null_evap_EF_simplified_rv}
		\begin{split}
			\ev{\T_{vr}}{U} &= -\frac{2}{f_0} \left( \ev{\T_{uu}}{U} + \ev{\T_{uv}}{U} \right) = -\frac{\varepsilon}{64\pi G_N r_0^2} \left( 1 + \frac{r_0}{r} + \frac{r_0^2}{r^2} - \frac{7r_0^3}{r^3} \right) , \\
			\ev{\T_{vv}}{U} &= \ev{\T_{uu}}{U} + 2 \ev{\T_{uv}}{U} + \ev{\T_{vv}}{U}_{(u, v)} = \frac{\varepsilon}{128\pi G_N r_0^2} \left( 1 - \frac{16 r_0^3}{r^3} + \frac{14 r_0^4}{r^4} \right).
		\end{split}
	\end{equation}
	
	The quantum correction to the metric is then easily solved for using \eqref{eq:evap_rr_eom_sc}-\eqref{eq:evap_vr_eom_sc} as
	\begin{equation} \label{eq:m_h_soln_evap}
		\begin{split}
			m(r, v) &= \frac{1}{8\lambda^2 r_0} \left( -\frac{7}{2} \frac{r_0^2}{r^2} + \frac{r_0}{r} - \log r - \frac{r}{r_0} \right) + D_3(v),\\
			h(r, v) &= \frac{1}{8\lambda^2 r_0^2} \left( -\frac{3}{2} \frac{r_0^2}{r^2} - 2 \frac{r_0}{r} + \log r \right) + D_4(v).
		\end{split}
	\end{equation}
	Finally we use \eqref{eq:evap_vv_eom_sc} and get
	\begin{equation} \label{eq:D3_relation}
		\dot{D}_3(v) = \frac{1}{16\lambda^2 r_0^2} \implies D_3(v) = D_3^{(\text{reg})} + \frac{v - v_0}{16 \lambda^2 r_0^2},
	\end{equation}
	where $D_3^{(\text{reg})}$ is the large distance regularization part that plays a similar role to $D_1$ and is given by
	\begin{equation} \label{eq:D3_regularization}
		D_3^{(\text{reg})} = \frac{1}{8\lambda^2 r_0} \left( \frac{L}{r_0} + \log L \right).
	\end{equation}
	
	The function $D_4(v)$ can be fixed by requiring that $h(r, v) \to 0$ at $r = L$ as $L \to \infty$ at fixed $v$, for every $v$. Hence the quantum-corrected part of the metric coefficients to the linear order is
		\begin{equation} \label{eq:m_h_soln_evap_final}
			\begin{split}
				m(r, v) &= \frac{1}{8\lambda^2 r_0} \left( -\frac{7}{2} \frac{r_0^2}{r^2} + \frac{r_0}{r} - \log \frac{r}{L} - \frac{r}{r_0} + \frac{L}{r_0} + \frac{v - v_0}{2 r_0} \right),\\
				h(r, v) &= \frac{1}{8\lambda^2 r_0^2} \left( -\frac{3}{2} \frac{r_0^2}{r^2} - 2 \frac{r_0}{r} + \log \frac{r}{L} \right).
			\end{split}
		\end{equation}

	We can then locate the apparent and event horizons from \eqref{eq:AH} and \eqref{eq:EH_perturb} at
	\begin{equation} \label{eq:event_apparent_hors_appndx}
		\begin{split}
			r_h^{\text{AH}}(v) &= r_0 + \frac{\varepsilon}{8\lambda^2 r_0} \left( \frac{7}{2} - \frac{L}{r_0} + \log \frac{r_0}{L} - \frac{(v - v_0)}{2r_0} \right), \\
			r_h(v) &= r_0 + \frac{\varepsilon}{8\lambda^2 r_0} \left( \frac{5}{2} - \frac{L}{r_0} + \log \frac{r_0}{L} - \frac{(v - v_0)}{2r_0} \right) = r_h^{\text{AH}}(v) - \frac{\varepsilon}{8\lambda^2 r_0}\, .
		\end{split}
	\end{equation}
	The event horizon is always located within the apparent horizon, which is to be expected. Finally, the dynamical surface gravity, which is also related to the local temperature of the evaporating black hole \cite{hayward2008localtemp} is given by
	\begin{equation} \label{eq:surface_grav_sc_evap}
		\kappa(v) = \frac{1}{2r_0} + \frac{\varepsilon}{16\lambda^2 r_0^4} \left( L - 3r_0 + \frac{v - v_0}{2} \right) = \kappa_0 + O(\varepsilon).
	\end{equation}

	\bibliographystyle{JHEP}
	\bibliography{Waldref}
\end{document}